\documentclass[superscriptaddress,a4paper,showpacs,nofootinbib]{revtex4}

\usepackage[normalem]{ulem}
\usepackage{amsmath}
\usepackage{graphicx,color}
\usepackage{array}

\DeclareMathOperator{\prob}{\mathrm{Proba}}
\DeclareMathOperator{\variance}{\mathrm{Variance}}

\begin{document}

\title{Effect of selection on ancestry: an exactly soluble case\\
and its phenomenological generalization}

\date{\today}

\author{\'E. Brunet}
\author{B. Derrida}
\affiliation{Laboratoire de Physique Statistique, \'Ecole Normale
Sup\'erieure,
24 rue Lhomond, 75231 Paris cedex 05, France}
\author{A.~H. Mueller}
\affiliation{Department of Physics, Columbia University,
New York, NY 10027, USA}
\author{S. Munier}
\affiliation{Centre de Physique Th{\'e}orique, 
{\'E}cole Polytechnique, CNRS, 91128~Palaiseau, France}

\begin{abstract}
We consider a family of models describing the evolution under selection
of a
population whose dynamics can be related to the propagation of noisy
traveling waves. For one particular model, that we shall call the
exponential model, the
properties of the traveling wave front can be calculated exactly, as well
as the statistics of the genealogy of the population. One striking result
is that, for this particular model, the genealogical trees have the same
statistics as the trees of replicas in the Parisi mean-field theory of
spin glasses. We also find that in the exponential model, the coalescence times along these trees grow like
the logarithm of the population size. A phenomenological picture of the
propagation of wave fronts that we introduced in a previous work, as
well as our numerical
data, suggest that these statistics remain valid for a larger class of
models, while the coalescence times
grow like the cube of the logarithm of the population size.
\end{abstract}

\maketitle


\section{Introduction}

It has been recognized for a long time that there is a strong analogy
between neo-darwinian evolution and statistical mechanics
\cite{Peliti.97}. For an evolving population, there is an ongoing
competition between the mutations which make individuals explore larger
and larger regions of genome space and selection which tends to
concentrate them at the optimal fitness genomes. This is very similar to
the competition
between the energy and the entropy in statistical mechanics.

In the simplest models of evolution, one associates to each individual
\cite{Tsimring.96,Kessler.97} (or to each species \cite{BakSneppen.93}) a
single number which represents how fit this individual is to its
environment. This fitness is transmitted to the offspring, up to small
variations due to mutations. A higher fitness usually means a larger
number of offspring
\cite{Tsimring.96,Kessler.97,KlosterTang.04,Kloster.05,Snyder.03,BDMM.06,BDMM2.06}.
If the size of the population is limited by the available resources,
survivors are chosen at random among all the offspring. This leads in the
long term to a selection effect: the descendants of individuals with low
fitness are eliminated whereas the offspring of the individuals with high
fitness tend to overrun the whole population.

\begin{figure}[!ht]
\begin{center}
 \includegraphics[width=12cm]{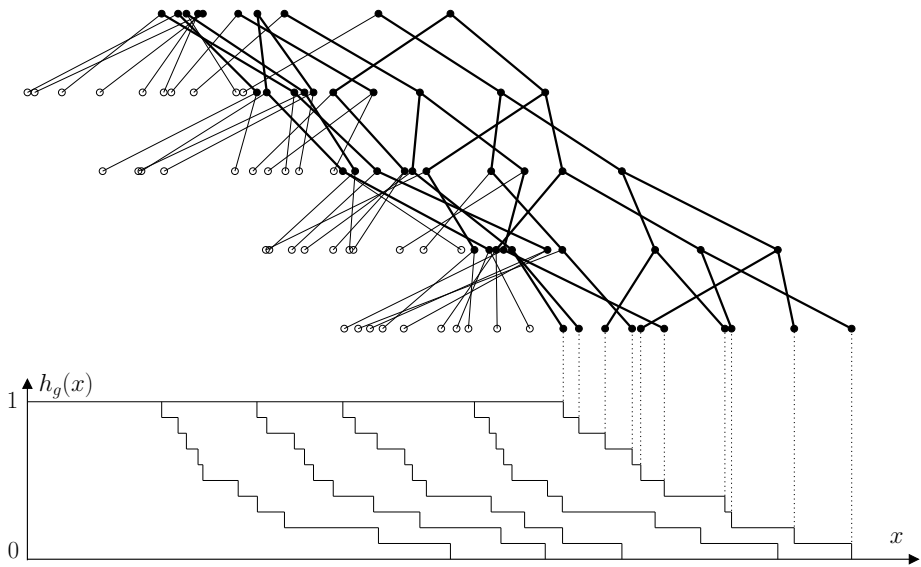}
\end{center}
\caption{\label{fig1}Numerical simulation of the evolution of model A,
with $k=2$ and $\rho(\epsilon)$ uniform between $-\frac12$ and $\frac12$
for $N=10$. \textit{Upper plot:} The filiation between each individual
and its two offspring is shown. The individuals eliminated at each
generation are shown by white circles, the surviving ones are shown by
black disks. \textit{Lower plot:} The noisy traveling wave front
$h_g(x)$, constructed as in (\ref{delta1}), is shown for the five
generations of the upper plot. }
\end{figure}

\begin{figure}[!ht]\centering
\includegraphics[width=10cm]{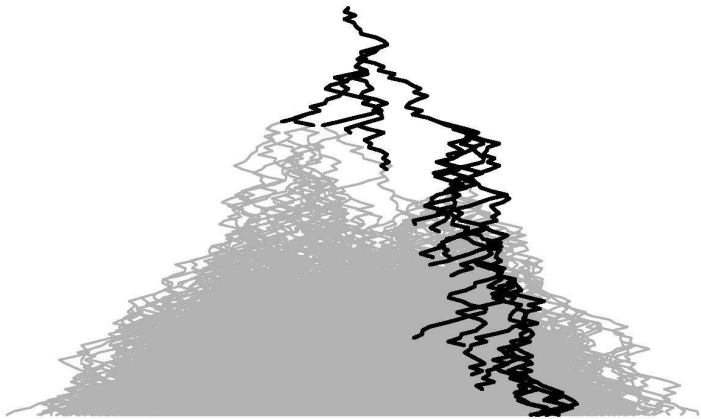}
\caption{A branching process for which the size $N$ of the population is
limited to 5. Each time the number of walks reaches 6, the leftmost walk
is eliminated. Time goes downwards and the horizontal direction
represents space. The actual population is represented in black, while
the grey lines represent what the population would be for infinite $N$
(\textit{i.e.} in the absence of selection).}
\label{A5}
\end{figure}

Our focus in this paper is a class of such models
\cite{KlosterTang.04,Kloster.05,Snyder.03,BDMM.06,BDMM2.06} describing
the evolution of a population of fixed size $N$ under asexual
reproduction. The $i$-th individual is characterized by a single real
number, $x_i(g)$, which represents its adequacy to the environment. (This
$x_i(g)$ plays a role similar to fitness in the sense that offspring with
higher $x_i(g)$ will be selected; in the following, we shall simply call
it the position of the individual.) At a generation $g$, the population
is thus represented by a set of $N$ real numbers $x_i(g)$ for $1\le i\le
N$. At each new generation, all individuals disappear and are replaced by
some of their offspring: the $j$-th descendant of individual $i$ has
position $x_i(g)+ \epsilon_{i,j}(g)$ where $  \epsilon_{i,j}(g)$
represents the effect of mutations from generation $g$ to generation
$g+1$. Then comes the selection step: at generation $g+1$, one only keeps
the $N$ rightmost offspring among the descendants of all individuals at
generation $g$. One may consider two particular variants of this model:

\noindent
\textit{Model A}:  each individual  has a fixed number $k$ of offspring and
all the $ \epsilon_{i,j}(g)$ are independently distributed according to
a given distribution $\rho(\epsilon)$. For example,
$\rho(\epsilon)$ may be the
uniform distribution between 0 and 1. A realization of such an evolution
is shown in figure~\ref{fig1}. Another example would be $N$
branching random walks where the size of population is kept constant
by eliminating the leftmost walk each time a branching event occurs.
A visual representation of this latter example is shown in figure~(\ref{A5}).

\noindent
\textit{Model B}: each individual has infinitely many offspring:
the $\epsilon_{i,j}(g)$  are
distributed according to a Poisson process of density
$\psi(\epsilon)$ (this means that, with probabiliy $\psi(\epsilon)
d\epsilon$, there  is one offspring
of individual $i$ with position between $x_i(g) + \epsilon$ and $
x_i(g) + \epsilon+ d \epsilon $).
The density $\psi(\epsilon)$ is  \textit{a priori} arbitrary. The only constraints we
 impose 
are that $\psi(\epsilon)$
decays fast enough, when $\epsilon$ increases,
for the position not
to diverge after one generation,
and that $\int_{- \infty}^\infty \psi(\epsilon) d \epsilon =
\infty$, for the survival probability to be 1.
(This latter constraint implies in fact that each
individual $i$
has infinitely many offspring before the selection step.)

As discussed in section \ref{sec:FKPP}, these models are related to
noisy traveling wave equations, of the Fisher-KPP type
\cite{Fisher.37,KPP.37,vanSaarloos.03}, which appear in many contexts: disordered systems
\cite{DerridaSpohn.88,BrunetDerrida.04}, reaction-diffusion
\cite{Breuer.95,DoeringMuellerSmereka.03,Lemarchand.99,Moro.01},
fragmentation \cite{KrapivskyMajumdar.00} or QCD
\cite{MunierPeschanski.03,IancuMuellerMunier.05,MuellerShoshi.04}.
A number of recent works
\cite{MuellerSowers.95,Mai.96,Kessler.98,Moro.01,Panja.03,Escudero.04,Moro.04,Moro2.04,ConlonDoering.05,BDMM.06}
focused on the fluctuations of the position
these fronts, and this will allow us to predict how the
fitness of the population evolves with the number of generations.

Another interesting aspect of these models with stochastic evolution is
their genealogy \cite{BDMM2.06}: one can associate to any group of
individuals at a given generation its genealogical tree. One can then
study how this tree fluctuates, and in particular what is the number of
generations needed to reach their most recent common ancestor. 
The relationship between noisy traveling waves and
genealogies is the main purpose of the present paper.

While the models we consider here are difficult
to solve for arbitrary $\rho(\epsilon)$ and $\psi(\epsilon)$, one
particular case of model~B, with $\psi(\epsilon) = e^{-\epsilon}$,
turns out to be analytically solvable both for
the statistics of the position of the population
 and for the properties of the genealogical trees.
We shall call this case the ``exponential model'' and 
present its solution in section~\ref{exact}.

As explained at the end of section~\ref{sec:FKPP}, the exponential model is
however non generic in the sense that it does not behave like a
Fisher-KPP front. The generic case (which behaves like a noisy
Fisher-KPP equation but that we are not
able to solve) and the exponential model can however
be both described by a similar
phenomenological theory \cite{BDMM.06}, that we develop in section~\ref{pheno}.
As a consequence, we argue that both the generic case and the
exponential model have the same cumulants for the position of the
front (up to a change of scale), and that the genealogical trees
have the same statistics in both models (up to a change of time scale).
Numerical results, presented in section~\ref{sec:numerical}, support
these claims.

\section{The link with noisy Fisher-KPP fronts}
\label{sec:FKPP}

Our models are nothing but stochastic models for the evolution of the
positions of $N$ individuals along the real axis.
These positions form a cloud which
does not spread: if an individual happens to fall
far behind the cloud, it will
have no surviving offspring, whereas the descendants of an
individual far ahead of the cloud
grow till they replace the whole population. With this
picture in mind, it makes sense to describe the population by a
front. Let $N h_g(x)$ be the number of individuals with a position larger
than $x$:
\begin{equation}
h_g(x)={1\over N}\int_x^{\infty}dz\,\sum_{i=1}^N\delta\big(z-x_i(g)\big).
\label{delta1}
\end{equation}
Clearly, $h_g(x)$ is a decreasing function with $h_g(-\infty)=1$
and $h_g(+\infty)=0$. 
In this section, we write the noisy equation which governs the evolution
of this front.

Let $N h_{g+1}^*(x)$ be the number of offspring on the
right of $x$ at generation $g+1$ \emph{before the selection step}. (So,
for instance, $h_{g+1}^*(-\infty)$ is $k$ in model~A and $\infty$ in
model~B). Once $h_{g+1}^*(x)$ is known, the selection step to
get $h_{g+1}(x)$ is simply:
\begin{equation}
h_{g+1}(x)=\min\left[1,h_{g+1}^*(x)\right].
\label{gg*}
\end{equation} 

Let us write the average and variance of 
$h_{g+1}^*(x)$ for both models.

\subsection{Statistics of $h_{g+1}^*(x)$ for model~A}

In model~$A$, one can write
\begin{equation}
N h_{g+1}^*(x) = \sum_{i=1}^N n_{g+1}^{(i)}(x),
\label{mA1}
\end{equation}
where $n_{g+1}^{(i)}(x)$ is the total number of offspring before
selection of the $i$-th individual of generation $g$  which fall on the
right of $x$.  
The probability that an offspring of $i$ falls on the right of $x$ is
$\int_{x}^{\infty}d\epsilon\,\rho(\epsilon-x_i)$ and, as the $k$ offspring of
$x_i(g)$ are independent, $n_{g+1}^{(i)}(x)$ has a
binomial distribution.
The average and variance are therefore given by
\begin{equation}
\overline{n_{g+1}^{(i)}(x)}
=k\int_x^{\infty}d\epsilon\,\rho\big(\epsilon-x_i(g)\big),\qquad
\variance\Big({n_{g+1}^{(i)}(x)}\Big)
=k\int_x^{\infty}d\epsilon\,\rho\big(\epsilon-x_i(g)\big)
	\left(1-\int_x^{\infty}d\epsilon\,
\rho\big(\epsilon-x_i(g)\big)\right).
\end{equation}
As the variables $n_{g+1}^{(i)}(x)$ are uncorrelated, the average
and variance of $N h_{g+1}^*(x)$ are simply from (\ref{mA1})
the sums over $i$ of the averages and variances of the $n_{g+1}^{(i)}(x)$.
For the average, one has
\begin{equation}
N \overline{h_{g+1}^*(x)} = k\int_x^{\infty}d\epsilon\,
\sum_i\rho\big(\epsilon-x_i(g)\big)
=-k\int_x^{\infty}d\epsilon\,
\int dz\,\rho(\epsilon-z)Nh'_g(z),
\label{deltaA}
\end{equation}
Where we used, from (\ref{delta1}),
\begin{equation}
\sum_{i=1}^N\delta\big(x-x_i(g)\big)=-Nh'_g(x).
\label{delta2}
\end{equation}
Simplifying, and doing the same transformation
for the variance, one finally gets
\begin{equation}
\overline{ h_{g+1}^*(x)}=k\int d\epsilon\,h_g(x-\epsilon)\rho(\epsilon)
,\quad
\variance\left(h_{g+1}^*(x)\right)={k\over N}\int
d\epsilon\,h_g(x-\epsilon)\rho(\epsilon)
\left[1-2\int_\epsilon^{\infty}dz\,\rho(z)\right]
\ \text{for model A}.
\label{mA2}
\end{equation}
(Note that these average and variance are obtained for a given $h_g(x)$:
they are not computed for the whole history.)

\subsection{Statistics of $h_{g+1}^*(x)$ for model~B}

In model~B, before the selection step, an individual at position $x_i(g)$
has infinitely many offspring given by a Poisson process of density
$\psi\big(x-x_i(g)\big)$. As Poisson processes are additive,
the whole population (before selection) at
generation $g+1$ is also given by a Poisson process of density~$\Psi(x)$ with
\begin{equation}
\Psi(x)=\psi\big(x-x_1(g)\big)+\cdots+\psi\big(x-x_N(g)\big).
\label{bigfish}
\end{equation}
The number of individuals on the right of $x$ is therefore a
Poisson random number of average $\int_x^{\infty}d\epsilon\,
\Psi(\epsilon)$, thus
\begin{equation}
\overline{N h_{g+1}^*(x)}=\variance\left(N h_{g+1}^*(x)\right)
=\int_x^{\infty}d\epsilon\, \Psi(\epsilon).
\label{mB1}
\end{equation}
One can rewrite $\Psi(\epsilon)$ using the same trick as in
(\ref{delta2}) and (\ref{deltaA}). One finally gets
\begin{equation}
\overline{ h_{g+1}^*(x)}=\int d\epsilon\,h_g(x-\epsilon)\psi(\epsilon)
\qquad\text{and}
\qquad
\variance\left(h_{g+1}^*(x)\right)={1\over N}\int
d\epsilon\,h_g(x-\epsilon)\psi(\epsilon)
\qquad\text{for model B}.
\label{mB2}
\end{equation}

\subsection{Front equations for both models and comparaison to
Fisher-KPP fronts}

Comparing (\ref{mB2}) and (\ref{mA2}), one sees that one can
write, for both models
\begin{equation}
h_{g+1}^*(x)=\overline{
h_{g+1}^*(x)}+\eta_g(x)\sqrt{\variance\big({h_{g+1}^*(x)}\big)},
\end{equation}
where $\eta_g(x)$ is a noise with $\overline{\eta_g(x)}=0$
and $\variance\big(\eta_g(x)\big)=1$. 
Using (\ref{gg*}) one finally gets
\begin{subequations}
\label{hgx}
\renewcommand{\theequation}{\theparentequation\,\Alph{equation}}
\begin{align}
h_{g+1}(x)&=\min\left[1,k\int
d\epsilon\,h_g(x-\epsilon)\rho(\epsilon)+
{\eta_g(x)\over\sqrt N}\sqrt{{k}\int
d\epsilon\,h_g(x-\epsilon)\rho(\epsilon)
\left(1-2\int_\epsilon^{\infty}dz\,\rho(z)\right)}\,\right]
\label{hgxA}
&\text{for model A},\\
h_{g+1}(x)&=\min\left[1,\int
d\epsilon\,h_g(x-\epsilon)\psi(\epsilon)+
{\eta_g(x)\over\sqrt N}\sqrt{\int
d\epsilon\,h_g(x-\epsilon)\psi(\epsilon)}\,\right]
&\text{for model B}.\label{hgxB}
\end{align}
\end{subequations}
The precise distribution of $\eta_g(x)$ depends on $N$ and on
the choice of the model. Far from both tips of the front,
this distribution is Gaussian. At the tip, however, where $h_{g}(x)$ is
of order $1/N$, both $\overline{h_{g}(x)}$ and its variance are
comparable and the noise cannot be approximated by a Gaussian. (This is
because the number of individuals is small and the discrete character of
$h_g(x)$ cannot be forgotten anymore.)
Furthermore, the noise
is correlated in space but uncorrelated for different $g$.

Thus, the precise expression of the noise $\eta_g(x)$
is rather complicated, but its variance is 1, so that
the amplitude of the whole noise term in (\ref{hgx})  
decays as $1/\sqrt{N}$ as $N$ becomes large.

Equations (\ref{hgx}) are very similar to the noisy Fisher-KPP equation:
\begin{equation}
{\partial h_g(x)\over\partial g}={\partial^2 h_g(x)\over\partial x^2}
+{h_g(x) -h_g(x)^2} +{\eta_g(x)\over\sqrt{N}} \sqrt{h_g(x)
-h_g(x)^2},
\label{F-KPP}
\end{equation}
where $\eta_g(x)$ is a Gaussian noise with $\overline{\eta_g(x)}=0$
and
$\overline{\eta_g(x)\eta_{g'}(x')}=\delta(g-g')\delta(x-x')$. The noisy
Fisher-KPP equation appears as a dual equation for the branching process
$A\to2A$ (rate~1) and $2A\to A$ (rate~$1/N$) or, more simply, is an
approximate equation valid for large $N$ describing the fraction of
$A$ in the chemical reaction $A+B\to2A$ when the concentration of
reactants is of order $N$
\cite{DoeringMuellerSmereka.03,PechenikLevine.99,Lemarchand.95}

Comparing (\ref{hgx}) and (\ref{F-KPP}),
the convolution of $h_g(x)$ by $k\rho(\epsilon)$ or $\psi(\epsilon)$ in
(\ref{hgx}) spreads the front in the same way as the diffusion term
in (\ref{F-KPP}). The same convolution induces the growth,
similarly to the linear $h_g(x)$ term in (\ref{F-KPP}),
as $k\phi(\epsilon)$ and $\psi(\epsilon)$  both have an integral larger than
1. Thus, the
fixed point $h_g(x)=0$ is unstable. To
balance the indefinite growth of $h_g(x)$, both (\ref{hgx}) and
(\ref{F-KPP}) have a saturation mechanism (respectively the $\min(1,\dots)$
and the $-h_g(x)^2$ term) which makes  $h_g(x)=1$ a stable fixed point.
So, ignoring the noise terms ($N\to\infty$),  both
(\ref{hgx}) and (\ref{F-KPP}) describe a front which propagates
from a stable phase $h_g(x)=1$ into an unstable phase $h_g(x)=0$.
Finally, the noise terms in (\ref{hgx}) and (\ref{F-KPP}) have a similar
amplitude of the order of $\sqrt{h_g(x)/N}$ in the unstable region
$h_g(x)\ll1$.

It is clear from the definitions of our models that the average velocity
of the front is an increasing function of $N$. We first consider the
limiting case $N\to\infty$, which is equivalent to removing the noise
term ($\eta_g=0$) from (\ref{hgx}) and (\ref{F-KPP}). 
To determine \cite{vanSaarloos.03} the velocity of such traveling
wave equations, it is usually sufficient to consider the linearized
equation in the unstable region $h_g(x)\ll1$ (where the saturation
mechanism can be neglected). Looking for solutions of the form
$h_g(x)\simeq \exp[-\gamma(x-v g)]$, one gets a relation between the
decay rate $\gamma$ and the velocity $v=v(\gamma)$ that reads
\begin{align}
v(\gamma)&={1\over\gamma}\ln\left[k\int
d\epsilon\,\rho(\epsilon)e ^{\gamma\epsilon}
\right]\quad\text{for model A},
&v(\gamma)&={1\over\gamma}\ln\left[\int
d\epsilon\,\psi(\epsilon)e ^{\gamma\epsilon}
\right]\quad\text{for model B}.\label{vg}
\end{align}
(For Fisher-KPP (\ref{F-KPP}), one has $ v(\gamma)=\gamma^{-1}+\gamma$.)

In many cases, when $v(\gamma)$ is finite over some range of $\gamma$
and reaches a minimal value $v(\gamma_0)$ for some finite positive decay
rate $\gamma_0$, the selected velocity of the front 
for a steep enough initial condition \cite{vanSaarloos.03} 
is this minimal velocity $v(\gamma_0)$.
For instance, for (\ref{F-KPP}), one
has $\gamma_0=1$ and the selected velocity is $v(\gamma_0)=2$.
Whenever this minimal velocity exists, we shall say that the model
is in the universality class of the Fisher-KPP equation (\ref{F-KPP}).
For finite $N$, \textit{i.e.} in presence of noise, there is a correction
to this velocity and the front diffuses. We shall recall \cite{BDMM.06} in
section~\ref{pheno} that for the generic Fisher-KPP case, the correction
to the velocity is of order $1/\ln^2N$ and that the diffusion constant is
of order $1/\ln^3N$.

There are however some choices of $\rho(\epsilon)$ or $\psi(\epsilon)$
for which $v(\gamma)$ is everywhere infinite or has no minimum.
An example  which we study in some detail in
section~\ref{exact} is model~B with $\psi(\epsilon)=e^{-\epsilon}$, for
which $v(\gamma)=\infty$ for all $\gamma$. We shall see among other
things that, in presence of noise, the velocity of that front diverges as
$\ln\ln N$ for large $N$ instead of converging to a finite value.

It has been known for a long time that traveling wave equations are
related to branching random walks \cite{McKean.75,Bramson.78}. This can
be seen by considering a single individual at the origin at generation~0
and by looking at the evolution of the probability $Q_g(x)$ that all
of its descendants at generation $g$ are on the left of $x$. In the
case of model~$B$ with $N=\infty$, one has
\begin{equation}
Q_{g+1}(x)=\prod_{y} [1-\psi(y)dy+\psi(y)dy\, Q_g(x-y)]=
\exp\left({\int dy\,\psi(y)(Q_g(x-y)-1)}\right).
\end{equation}
This equation describes the propagation of a front of the Fisher-KPP
type, but where the unstable fixed point is at $Q_g=1$ instead of $0$. For $Q_g$ close to 1,
one gets exponentially decaying traveling wave solutions of the form
$1-Q_g(x)\propto \exp[{-\gamma(x-vg)}]$, with $v=v(\gamma)$ given by
(\ref{vg}) for model~B. (A similar calculation for model~A leads to
$v(\gamma)$ given by~(\ref{vg}).)


\section{Exact results for the exponential model}
\label{exact}
In this section, we derive exact expressions (for large $N$) of the
velocity, diffusion constant and coalescence times for model~B with
$\psi(\epsilon)=e^{-\epsilon}$.
We first write some expressions valid for model~B with an arbitrary
density function $\psi(\epsilon)$, which we shall later apply to the
exponential model.

Before selection, the positions of the individuals at
generation $g+1$ are distributed according to
a Poisson process of density~$\Psi(x)$
defined in (\ref{bigfish}).
We now wish to know the distribution of the $N$ rightmost
individuals of this Poisson process. (\textit{i.e.} of the
offspring who survive the selection step.) We first consider the
probability that there are no offspring on the right of $x$. Clearly,
it is given by
\begin{equation}
\prod_{x<z<\infty}[1-\Psi(z)\,dz]
=\exp\left(-\int_x^\infty \Psi(z)\,dz\right).
\end{equation}
Then, the probability that the rightmost offspring at generation $g+1$
is in the interval
$[x_1,x_1+dx_1]$, and the second rightmost is in $[x_2,x_2+dx_2]$, up to
the $N+1$-st rightmost particle is
\begin{equation}
\Psi(x_{N+1})dx_{N+1}\,\Psi(x_N)dx_N\,\cdots\Psi(x_1)dx_1
\exp\left(-\int_{x_{N+1}}^\infty \Psi(z)\,dz\right)
\qquad\text{for $x_{N+1}<x_N<\cdots<x_1$}.
\label{prob_ordered}
\end{equation}
It will be more convenient
not to specify the ordering of the $N$ rightmost particles.
Then the probability that the $N+1$-st rightmost particle 
is in the interval $[x_{N+1},x_{N+1}+dx_{N+1}]$ (as before) and that
the $N$ rightmost particles are in the intervals $[x_k,x_k+dx_k]$
for $1\le k\le N$, with no constraint on the order of
$x_1,\dots,x_N$, becomes
\begin{equation}
{1\over N!}
\Psi(x_{N+1})dx_{N+1}\,\Psi(x_N)dx_N\,\cdots\Psi(x_1)dx_1
\exp\left(-\int_{x_{N+1}}^\infty \Psi(z)\,dz\right)
\qquad\text{when $x_{N+1}<x_k$ for $k=1,\dots,N$}.
\label{P_unordered}
\end{equation}

One obtains the probability
that the $N+1$-st rightmost particle is in the interval
$[x_{N+1},x_{N+1}+dx_{N+1}]$ 
 by integrating (\ref{P_unordered}) over $x_1,\dots,x_N$:
\begin{equation}
{1\over N!}
\Psi(x_{N+1}) dx_{N+1}
\left[\int_{x_{N+1}}^{\infty}\Psi(x)\,dx\right]^N
\exp\left(-\int_{x_{N+1}}^\infty \Psi(z)\,dz\right).
\label{P_N+1}
\end{equation}
(As we imposed
$\int_{-\infty}^{+\infty}\psi(\epsilon)\,d\epsilon=\infty$ in the
definition of the model,
this distribution is normalized; see (8).)
Finally, the probability of $x_1,\dots,x_N$ given $x_{N+1}$ is
the ratio of (\ref{P_unordered}) by (\ref{P_N+1}).
One can see that, given the value of $x_{N+1}$,
the distributions of $x_1(g+1),\dots,x_N(g+1)$ are independent and one
gets that, given $x_{N+1}$,
each of the $N$ rightmost particles
is in $[x,x+dx]$ with probability
\begin{equation}
{\Psi(x)\,dx\over\int_{x_{N+1}}^{\infty}\Psi(x)\,dx}
\qquad\text{for $x_{N+1}<x$}.
\label{x_indep}
\end{equation}

Therefore, to generate the whole population after selection at generation
$g+1$, one needs to calculate the density $\Psi(x)$ according
to (\ref{bigfish}), then to choose the position of the $N+1$-st rightmost
particle according to (\ref{P_N+1}) and, finally, to
generate \emph{independently} the $N$ rightmost particles
$x_1(g+1),\dots,x_N(g+1)$ with the distribution (\ref{x_indep}). Note that
the $N+1$-st particle is not selected and is therefore eliminated after
the $N$ rightmost particles have been generated.
This procedure is valid for any $\psi(\epsilon)$, but is in
general complicated because (\ref{bigfish}) is not easy to handle
analytically.

\subsection{Statistics of the position of the front in the exponential
model}

In the exponential model $\psi(\epsilon)=e^{-\epsilon}$, however,
everything becomes simpler: the Poisson process (\ref{bigfish}) becomes
\begin{equation}
\Psi_\text{exp}(x)=e^{-(x-X_g)}\quad\text{with
$X_g=\ln\left(e^{x_1(g)+x_2(g)+\cdots+x_N(g)}\right)$},
\label{defXg}
\end{equation}
which means that the offspring of the whole population is distributed as
if they were
the offspring of a single effective individual located at position $X_g$.
The distribution of the $N+1$-st rightmost particle (\ref{P_N+1}) becomes
\begin{equation}
x_{N+1}=X_g+z\qquad\text{with }\prob(z)=
{1\over
N!}\exp\left[-(N+1)z-e^{-z}\right],
\label{def_z}
\end{equation}
and, once $x_{N+1}$ has been chosen, the distribution (\ref{x_indep}) of
the $x_k(g+1)$ for $k=1,\dots,N$ becomes :
\begin{equation}
x_k(g+1)=x_{N+1}+y_k\qquad\text{with }\prob(y_k)=
e^{-y_k}\qquad\text{for $y_k>0$}.
\label{def_yk}
\end{equation}

We now recall the calculation of the statistics of the
position of the front \cite{BDMM2.06} which was done
for a similar model 
in~\cite{BrunetDerrida.04}, because we shall use later the same approach 
to calculate the statistics
of the genealogical trees.

There are many ways of defining the
position of the front
at a given generation $g$. One could consider the position of
its center of mass, or the position of the rightmost or leftmost
individual, or actually, any function of the positions $x_k(g)$ such that
a global shift of all the $x_k(g)$ leads to the same shift in
the position of the front. Because the front does not spread, 
the difference
between two such definitions of the position does not grow with time so
that, in the limit $g\rightarrow\infty$, all these definitions lead to
the same
velocity, diffusion constant and higher cumulants.

For the exponential model, it is convenient to  use $X_g$, defined in
(\ref{defXg}), as the position of the
front. Indeed, one can write
\begin{equation}
\Delta X_g = X_{g+1}-X_g = z +\ln\left(e^{y_1}
+e^{y_2}+\cdots+e^{y_N}\right),
\label{shifts}
\end{equation}
where the definitions and probability distributions of $z$ and $y_k$ are
given in (\ref{def_z}) and (\ref{def_yk}).
From (\ref{shifts}), the shifts $\Delta X_g$ are
uncorrelated random
variables, and the average velocity $v_N$ and diffusion constant
$D_N$ of the front are given by
\begin{equation}
v_N=\langle \Delta X_g \rangle,\qquad
 D_N = \langle \Delta X_g^2 \rangle - \langle \Delta X_g \rangle^2.
\end{equation}
More generally, all cumulants of the front position at a long 
time $g$ are simply $g$ times the cumulants of $\Delta X_g$.
To compute theses cumulants, we evaluate the generating function
$G(\beta)$ defined as
\begin{equation}
e ^{G(\beta)}=
\left\langle e^{-\beta\Delta X_g}\right\rangle
=\int dz\, \prob(z) e ^{-\beta z}
\int dy_{1}\,\prob(y_1)\cdots
\int dy_{N}\,\prob(y_N)
\left(e^{ y_1}+\cdots+e^{ y_N}\right)^{-\beta},
\label{G}
\end{equation}
and one obtains the cumulants by doing a small $\beta$ expansion:
\begin{equation}
G(\beta)=\sum_{n\geq 1}\frac{(-\beta)^n}{n!}
{\langle \Delta X_g^n\rangle_c}.
\label{defcum}
\end{equation}

Using (\ref{def_z}), the integral over $z$ is easy:
\begin{equation}
\int dz\, \prob(z)
e ^{-\beta z}
=\frac{1}{ N!}
\int dz\, \exp\left[{-(\beta+N+1)z-e^{- z}}\right]
=\frac{\Gamma(N+1+{\beta})}{\Gamma(N+1)}.
\label{intyN}
\end{equation}
To calculate the integrals over $y_i$ in (\ref{G}), one can use the
representation (valid for $\beta>0$)
\begin{equation}
Z^{-\beta}=\frac{1}{\Gamma(\beta)}
\int_0^{+\infty}d\lambda\,\lambda^{\beta-1}e^{-\lambda Z}
\label{Gammarep}
\end{equation}
with $Z=e ^{ y_1}+\cdots+e ^{ y_N}$.
This leads to the 
factorization of the integrals over $y_1,\cdots,y_N$. Replacing
$\prob(y_k)$ by its explicit expression from (\ref{def_yk}), one gets
for $\beta>0$ (a similar calculation can be made for $\beta>-1$)
\begin{equation}
e ^{G(\beta)}=
\frac{\Gamma(N+1+{\beta})}
{\Gamma(N+1)\Gamma({\beta})}
\int_0^{+\infty}{d\lambda}\,{\lambda^{{\beta-1}}}
I_0(\lambda)^{N},
\label{Gint}
\end{equation} 
where
\begin{equation}
I_0(\lambda)=\int_0^{+\infty}dy\, e^{- y-\lambda e ^{ y}}.
\label{defI0}
\end{equation}
One can rewrite $I_0(\lambda)$ in several ways:
\begin{equation}
\begin{aligned}
I_0(\lambda)&={\lambda}\int_{\lambda}^{+\infty}\frac{du}{u ^2} e ^{-u}
=1+\lambda(\ln\lambda+\gamma_E-1)+[e^{-\lambda}-(1-\lambda)]
-\lambda\int_0^{\lambda}du\,\frac{1-e^{-u}}{u},\\
&=1+\lambda(\ln\lambda+\gamma_E -1)
-\sum_{k=0}^{+\infty}
\frac{(-1)^k}{(k+1)(k+2)!}\lambda^{k+2},
\end{aligned}
\label{expansionI0}
\end{equation}
where $\gamma_E=-\Gamma'(1)$ is the Euler constant.

As $I_0(\lambda)$ is a monotonous  decreasing function, the integral
(\ref{Gint}) is dominated by $\lambda$ close to 0. In fact, using
(\ref{Gint}), one
can check that the range of values of $\lambda$ which dominate
(\ref{Gint}) is of the order of $1/[N\ln N]$. Indeed, if one makes the
change of variables
\begin{equation}\mu=\lambda N \ln N,\label{mulambda}\end{equation}
one gets $I_0(\lambda)^N$ 
for values of $\mu$ of order 1:
\begin{equation}
\begin{aligned}\relax
[I_0(\lambda)]^N&\simeq\exp\left[
N\lambda(\ln\lambda+\gamma_E-1)\right],\\
&\simeq\exp\left[{\mu\over\ln N}(\ln\mu-\ln N-\ln\ln
N+\gamma_E-1)\right],\\
&\simeq e^{-\mu}\left(1+\mu{\ln\mu-\ln\ln N+\gamma_E-1\over\ln N}+
{1\over2}\left[\mu{\ln\mu-\ln\ln N+\gamma_E-1\over\ln
N}\right]^2+\cdots\right),
\end{aligned}
\label{I0^N}
\end{equation}
where terms of order $1/N$ have been dropped. Replacing this expression
into (\ref{Gint}) and using
\begin{equation}
\int_0^{\infty}d\mu\,\mu^{x-1}e^{-\mu}(\ln\mu)^k={d^k\over
dx^k}\Gamma(x),
\label{derGamma}
\end{equation}
one gets:
\begin{equation}
\begin{aligned}
e^{G(\beta)}&\simeq\frac{\Gamma(N+1+\beta)}{\Gamma(N+1)\Gamma(\beta)}
\,{1\over(N\ln N)^\beta}
\left[\Gamma(\beta)+{\Gamma'(\beta+1)+\Gamma(\beta+1)[-\ln\ln
N+\gamma_E-1]\over\ln N}+\cdots\right],\\
&\simeq\frac{\Gamma(N+1+\beta)}{\Gamma(N+1)}\,{1\over(N\ln N)^\beta}
\left[1+{\beta\over\ln N}\left({\Gamma'(\beta+1)\over\Gamma(\beta+1)}
-\ln\ln N+\gamma_E-1\right)+\cdots\right].
\end{aligned}
\label{eGfin}
\end{equation}
(The next order is obtained in
appendix~\ref{appA}.)
The Stirling formula allows to simplify the expression:
\begin{equation}
\frac{\Gamma(N+1+{\beta})}
{\Gamma(N+1)}\,{1\over N^\beta}=1 +{\cal O}\left(1\over N\right).
\label{stirling}
\end{equation}
Then, one gets from (\ref{eGfin})
the following expression for the generating function:
\begin{equation}
G(\beta)={-\beta}\ln {\ln N}
-\frac{\beta}{\ln N}
\left(\ln\ln N+1-\gamma_E-\frac{\Gamma^\prime(1+\beta)}{\Gamma(1+\beta)}\right)
+{o}\left(\frac{1}{\ln N}\right).
\label{Gbeta0}
\end{equation}
(This expression was obtained assuming $\beta>0$, but one can show
that it remains valid for $\beta>-1$ by using, instead of
(\ref{Gammarep}), a different representation of $Z^{-\beta}$.)
Now one simply reads off the expressions of the cumulants of the
position of the front 
by comparing the expansion of (\ref{Gbeta0}) in powers of $\beta$
and (\ref{defcum}):
\begin{equation}
\begin{split}
v_N = \frac{\left\langle X_g\right\rangle}{g}=\left\langle
\Delta X_g\right\rangle
&=\ln\ln N
+\frac{1}{\ln N}(\ln\ln N+1)+\cdots
\\
D_N =\frac{\left\langle X_g^2 \right\rangle_c}{g}
=\left\langle \Delta X_g^2\right\rangle_c
&=\frac{\pi ^{2}}{3\ln N}+\cdots\\
\frac{\left\langle X_g^n\right\rangle_c}{g} 
=\left\langle \Delta X_g^n\right\rangle_c
& = \frac{n!\zeta(n)}{\ln N}
=\frac{n!}{\ln N}\sum_{i\geq 1}\frac{1}{i^n}+\cdots,
\end{split}
\label{resleadingexpo}
\end{equation}
up to terms of order $\ln\ln N/\ln^2 N$ that are computed
in appendix~\ref{appA}.
The velocity 
$v_N$ diverges for large $N$, in contrast with models of the Fisher-KPP
class for which $v_N$ has a finite large~$N$ limit. Note that velocities
which become infinite in the large $N$ limit occur in other models of
evolution with selection \cite{Kessler.97}.


\subsection{Trees in the exponential model}
\label{sec:treesexp}

Let us now consider the ancestors of a group of
$p\geq 2$ individuals  chosen at random 
in the population (of size $N$).
 Looking at their genealogy, one observes a tree
which fluctuates with the choice of the $p$ individuals and which
is characterized by its shape and coalescence times.

For model~B with an arbitrary density $\psi(\epsilon)$, the probability of
finding,
at generation $g+1$ before selection, an offspring in
$[x,x+dx]$ is $\Psi(x)\,dx$ with $\Psi$ given by (\ref{bigfish}). On the
other hand, the probability of finding in $[x,x+dx]$
an  offspring of
$x_i(g)$ is, by definition, $\psi\big(x-x_i(g)\big)\,dx$. Therefore, given an
offspring at generation $g+1$ and position $x$, the probability that its
parent was the $i$-th individual (at position $x_i(g)$) is
\begin{equation}
W_i(x)={\psi\big(x-x_i(g)\big)\over \Psi(x)}.
\label{Wi_general}
\end{equation}
For general $\psi(\epsilon)$, these probabilities 
$W_i(x)$ depend on $x$, making the calculation of these coalescence times
difficult.
In the exponential model, however,
(\ref{Wi_general}) becomes
\begin{equation}
W_i=e^{x_i(g)-X_g}={e^{x_i(g)}\over e^{x_1(g)}+\cdots+ e^{x_N(g)}}= 
{e^{y_i}\over e^{y_1}+\cdots+ e^{y_N}},
\label{defWi}
\end{equation}
where the $y_k=x_k(g)-x_{N+1}(g)$
are the exponential variables of (\ref{def_yk}).
Therefore the $W_i$ do not depend on $x$.
It follows that the probability $q_p$ that $p$ individuals at generation
$g+1$ have the same ancestor at generation $g$ is simply
\begin{equation}
q_p= \left\langle\sum_{i=1}^N W_i ^p \right\rangle,
\label{defqp}
\end{equation}
where the average is over the $y_i$ of (\ref{defWi}).
After performing this average, all the terms
in the sum over $i$ become equal since the $y_i$ are
identically distributed.
Therefore
\begin{equation}
q_p=N\langle W_1^p\rangle=
N
\int_0^{+\infty}
dy_1\, e^ {- y_1}\cdots
\int_0^{+\infty}
dy_N\, e^ {- y_N}
e ^{p y_1}\left(
e^{ y_1}+\cdots+e ^{ y_N}
\right)^{-p}.
\end{equation}
Using the representation (\ref{Gammarep}),
one obtains
\begin{equation}
q_p=\frac{N}{(p-1)!}\int_0^{+\infty}d\lambda\,
\lambda ^{p-1}
I_p(\lambda)
I_0(\lambda)
^{N-1}
\label{qpexact}
\end{equation}
in terms of the function
$I_0(\lambda)$
introduced in (\ref{defI0})
and of its derivatives
\begin{equation}
I_p(\lambda)=\int_0^{+\infty}dy\, e ^{(p-1) y-\lambda 
e^{ y}}=
(-)^p {d^p\over d\lambda^p}I_0(\lambda)=
\lambda ^{1-p}\int_\lambda ^{+\infty}du\, u ^{p-2} e ^{-u}.
\label{defIp}
\end{equation}
For small $\lambda$  one has, to the leading order,
\begin{equation}
I_0(\lambda)\simeq 1+\lambda(\ln\lambda+\gamma_E-1),\qquad
I_1(\lambda)\simeq -(\ln\lambda+\gamma_E),\qquad
I_p(\lambda)\simeq {(p-2)!\over\lambda^{p-1}}\quad\text{for $p\ge2$}.
\label{Ipsmall}
\end{equation}

So far, (\ref{qpexact}) is an exact expression and valid for
arbitrary $N$.
From now on, we will work at leading order in $\ln N$,
leaving the extension to subleading orders to 
appendix~\ref{appA}.

As for the obtention of (\ref{Gbeta0}) from (\ref{Gint}), the integral
over $\lambda$ is dominated by the region where $\lambda$ is of order
$1/[N\ln N]$. Doing the same change of variable $\mu=\lambda N\ln N$,
one gets $I_0(\lambda)^N\simeq e^{-\mu}$ and, using
(\ref{Ipsmall}),
$\lambda^{p-1} I_p(\lambda)\simeq(p-2)!$. Therefore, we obtain for
$p\ge2$
\begin{equation}
q_p=\frac{1}{\ln N}\frac{1}{p-1}\ .
\label{qp}
\end{equation}
We see that for large $N$ the probability that $p$ branches merge is of
the same order for all $p$, in contrast to the neutral model
(\cite{Kingman2.82,Kingman.82} and appendix \ref{appC}) for which 
$q^p$ is of order $1/N^{p-1}$, so that 
$q_2\gg q_3\gg q_4\gg\cdots$.

To calculate the moments of the coalescence times, it is convenient to
introduce the probability $r_p(k)$ that $p$ randomly chosen
individuals at generation $g+1$
have exactly $k$ ancestors at generation $g$.
In one generation, at leading order in $N$, only
a single coalescence may occur among the $p$ individuals,
and (\ref{qp}) tells us that
the coalescence probability goes like $1/\ln N$ (any additionnal coalescence
at the same generation would in fact cost an additional power of $1/\ln
N$; see appendix \ref{appA}.)
Consequently, we just need that $p-k+1$ individuals coalesce to one
ancestor, say individual number~$i$
(the probability is $W_i^{p-k+1}$),
and that none of the other individuals have $i$
as an ancestor (probability $(1-W_i) ^{k-1}$).
Altogether, this reads\footnote{In the mathematical litterature, one
would rather use the transition rates $\lambda_{b,q}$
which give the probability that out of $b$ individuals, the only event
is the coalescence of the $q$ first individuals
\cite{Pitman.99,Schweinsberg.00}. Clearly,
$r_p(k)=\binom{p}{k-1}\lambda_{p,p-k+1}$.
All the $\lambda_{b,q}$ can be obtained through a measure
$\Lambda$ through $\lambda_{b,q}=\int_0^1x^{q-2}(1-x)^{b-q}\Lambda(dx)$.
The exponential case corresponds to a uniform measure $\Lambda$,
studied in \cite{BolthausenSznitman.98}.}
\begin{equation}
r_{p}(k)=
\left(\begin{matrix}p\\k\!-\!1
\end{matrix}\right)
\left\langle
\sum_{i=1}^N
W_i ^{p-k+1}
(1-W_i)^{k-1}
\right\rangle.
\end{equation}
The factor $(1-W_i)^{k-1}$
may be expanded and 
the average may be expressed with the help of the $q_p$
defined in (\ref{defqp}):
\begin{equation}
r_p(k)=\left(
\begin{matrix}p\\ k\!-\!1
\end{matrix}\right)
\sum_{j=0}^{k-1}
\left(\begin{matrix}k\!-\!1\\ j
\end{matrix}\right)
(-1)^{k-1-j}q_{p-j}.
\label{rpint}
\end{equation}
Replacing (\ref{qp}) in (\ref{rpint}), 
one gets after some algebra
\begin{equation}
r_p(k)=\frac{1}{\ln N}
\frac{p}{(p-k)(p-k+1)}\ ,
\label{rp}
\end{equation}
which holds for $k<p$.
The probability $r_p(p)$ that there is no coalescence at all among the
$p$ individuals (that is to say, that all $p$ have distinct
ancestors) has a simple expression, which is obtained from a
completeness
relation:
\begin{equation}
r_p(p)=1-\sum_{k=1}^{p-1}r_p(k)=1-\frac{p-1}{\ln N}\ .
\label{sp}
\end{equation}
The knowledge of the probabilities $r_p(k)$ in (\ref{rp}) and (\ref{sp})
allows one to determine (in the large $N$ limit) all the statistical
properties of the trees.

We introduce the probability $P_p(g)$ that $p$ individuals have their first
common ancestor a number of generations $g$ in the past.
For $p\ge2$, one may write a recursion for $P_p(g)$ in the form
\begin{equation}
P_p(g+1)=\sum_{k=2}^{p} r_p(k) P_k(g)+r_p(1)\delta_g^0.
\label{Ppint}
\end{equation}
Using~(\ref{rp}) and (\ref{sp}), this becomes
\begin{equation}
P_p(g+1)-P_p(g)
=-\frac{p-1}{\ln N} P_p(g)
+\sum_{k=2}^{p-1}\frac{1}{\ln N}
\frac{p}{(p-k)(p-k+1)} P_k(g)+r_p(1)\delta_g^0.
\end{equation}
In the large-$N$ limit, the number of generations $g$
over which the coalescence occurs
is typically
$\ln N\gg 1$ (since the coalescence probabilities scale like
$1/\ln N$). 
It is then natural
to introduce the rescaled variable $t=g/\ln N$
and the corresponding coalescence probability 
$R_p(t)\,dt=P_p(g)\,dg$.
In this new variable, the recursion becomes for $t>0$
\begin{equation}
\frac{dR_p(t)}{dt}=-(p-1)R_p(t)+\sum_{k=2}^{p-1}
\frac{p}{(p-k)(p-k+1)}R_k(t).
\label{diffPp}
\end{equation}
This equation may be solved by introducing the generating function
\begin{equation}
\Psi(\lambda,t)=\sum_{p \geq 2} \lambda^{p-1} R_p(t), 
\label{genPsi}
\end{equation}
which turns the summation over $k$ in (\ref{diffPp}) into
\begin{equation}
\frac{d \Psi}{dt}
=[ (1 -\lambda) \ln(1-\lambda)]
\frac{d \Psi}{d\lambda} - [ \ln(1-\lambda)] \Psi .
\end{equation}
The general solution (which can be obtained by the method of
characteristics) reads
\begin{equation}
\Psi(\lambda,t)=\frac{1}{1-\lambda}\phi(e^{-t}\ln(1-\lambda)),
\end{equation}
where $\phi$ is an arbitrary function.
The initial condition for (\ref{diffPp}) is the probability that
all $p$ individuals coalesce between times $0$ and $dt$ (see (\ref{qp})):
\begin{equation}
R_p(t\!=\!0)\,dt=q_p\times\frac{dg}{dt}\,dt=\frac{dt}{p-1},
\end{equation}
and thus, (\ref{genPsi}) becomes
\begin{equation}
\Psi(\lambda,t\!=\!0)= -\ln(1-\lambda).
\end{equation}
This leads to 
\begin{equation}
\Psi(\lambda,t)= \frac{d}{dt} (1-\lambda)^{e^{-t}-1}.
\label{Psit}
\end{equation}
The expansion of (\ref{Psit}) in powers of $\lambda$ using
\begin{equation}
(1-\lambda)^{-a}=\frac{1}{\Gamma(a)}
\sum_{p=0}^{+\infty}\frac{\Gamma(p+a)}{\Gamma(p+1)}\lambda^p.
\end{equation}
leads through (\ref{genPsi}) to
\begin{equation}
R_p(t) =\frac{1}{(p-1)!} \frac{d}{dt} \frac{\Gamma(p-e^{-t})}
{\Gamma(1-e^{-t})}=\frac{1}{(p-1)!} \frac{d}{dt}
\left[(1-e^{-t})(2-e^{-t})\cdots(p-1-e^{-t})\right],
\label{Pm}
\end{equation}
which is just a polynomial of order $p-1$ in the variable $e^{-t}$.
More explicitely, for the first values of $p$, one finds
\begin{equation}
\begin{split}
R_2(t)&= e^{-t}\ ,\ \  
R_3(t)= \frac{3}{2} e^{-t}  - e^{-2 t}\ ,\ \
R_4(t)= \frac{11}{6} e^{-t}  - 2 e^{-2 t} +\frac12 e^{-3 t}\ ,\ \ 
\ldots
\end{split}
\label{R_p}
\end{equation}
The average coalescence times (using (\ref{Pm})) are
\begin{equation}
\langle T_p\rangle=\sum_{g=0}^{\infty}g P_p(g)
=\ln N\int_0^{+\infty} dt\,t\,R_p(t)
=\ln
N\int_0^{\infty}dt\left[1-(1-e^{-t})
	\left(1-{e^{-t}\over2}\right)
	\left(1-{e^{-t}\over3}\right)\cdots
	\left(1-{e^{-t}\over p-1}\right)\right]
\label{calculTp}
\end{equation} 
and one gets
\begin{equation}
\begin{split}
\langle T_2 \rangle &= \ln N, \ \
\langle T_3 \rangle = \frac54 \langle T_2 \rangle,\ \ 
\langle T_4 \rangle = \frac{25}{18} \langle T_2 \rangle,\ \dots\\
\end{split}
\label{Tpexplicit}
\end{equation}
These expressions 
contrast with a neutral model of coalescence  with
no selection \cite{Kingman.82,TavareBGD.97} 
where at each generation one would choose the $N$ survivors at random
among all the offspring at generation $g+1$ (see
appendix~\ref{appC}):
\begin{equation}
\begin{split}
\langle T_2^\text{neutral} \rangle =  {\cal O}(N)\ ,\ \ 
\langle T_3^\text{neutral} \rangle = \frac43 \langle
T_2^\text{neutral} \rangle\ ,\ \
\langle T_4^\text{neutral} \rangle = \frac32 \langle
T_2^\text{neutral} \rangle\ ,\ \ \ldots\\
\end{split}
\label{Tneutral}
\end{equation}
(Table~\ref{treesneutral} compares the frequencies of the trees
in the cases with and without selection.)

\begin{table}[!ht]
\centering
\begin{tabular}{|m{11mm}|c|c|}
\hline
&Neutral case&Exponential model\\\hline
\includegraphics[width=11mm]{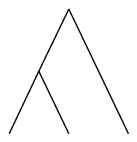}
&$\displaystyle1$&$\displaystyle{3\over4}$\\
\hline
\includegraphics[width=11mm]{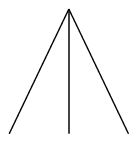}
&$\displaystyle0$&$\displaystyle{1\over4}$\\
\hline
\end{tabular}
\begin{tabular}{|m{11mm}|c|c|}
\hline
&Neutral case&Exponential model\\\hline
\includegraphics[width=11mm]{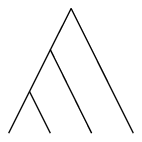} &$\displaystyle{2\over3}$&$\displaystyle{1\over3}$\\
\hline
\includegraphics[width=11mm]{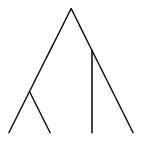} &$\displaystyle{1\over3}$&$\displaystyle{1\over6}$\\
\hline
\includegraphics[width=11mm]{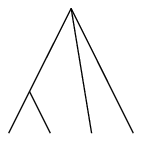} &$\displaystyle0$&$\displaystyle{1\over6}$\\
\hline
\includegraphics[width=11mm]{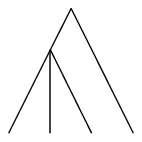} &$\displaystyle0$&$\displaystyle{2\over9}$\\
\hline
\includegraphics[width=11mm]{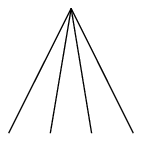} &$\displaystyle0$&$\displaystyle{1\over9}$\\
\hline
\end{tabular}\caption{Probabilities of observing each of the possible genealogical trees
for three and four individuals in the neutral case and in the exponential model}
\label{treesneutral}
\end{table}

As shown in appendix~\ref{appB}, the ratios~(\ref{Tpexplicit})
are on the other hand identical to those which would be computed
if the genealogical trees had the same statistical properties as
mean-field spin glasses \cite{Parisi.83,BolthausenSznitman.98}.

We also see that $\langle T_p\rangle$ in (\ref{Tpexplicit})
scales like $\ln N$ for any fixed
value of $p$, which means that on average, a given number of
individuals have their first common ancestor at of order $\ln N$
generations in the past. It is however interesting to note that 
for large $p$,
\begin{equation}
\langle T_p\rangle\simeq \ln N\times\ln\ln p
\label{largeNTN}
\end{equation} 
which is obtained by using, from (\ref{Pm}), $R_p(t)\simeq{d\over dt}
p^{-\exp({-t})}\simeq{d\over dt}
e^{-\exp[{-(t-\ln\ln p)}]}$ for large $p$; $R_p(t)$ becomes a Gumbel
distribution of width of order~1 centered at $\ln\ln p$.


\section{\label{pheno}Phenomenological extension to generic models}

The exponential model had the advantage of being exactly solvable, but as
already mentioned, it is non-generic because the velocity $v_N\to\infty$
as $N\rightarrow\infty$, in contrast to models of the Fisher-KPP type. We
do not know how to calculate directly the velocity $v_N$, diffusion
constant $D_N$ or the coalescence times of the generic Fisher-KPP case.
One can however use a phenomenological picture of front propagation
\cite{BDMM.06} and ancestry, which is fully consistent with exact
calculations in the case of the exponential model, and with numerical
simulations in the generic case.

\subsection{Picture of the propagation of fluctuating pulled fronts}
\label{phenoA}

Let us recall the phenomenological picture of front propagation
which emerged from \cite{BrunetDerrida.97,BDMM.06}.
In this picture, most of the time, the front evolves in a deterministic
way well reproduced by an equation obtained from (\ref{hgx}) by
removing the noise term, and by adding a cutoff which takes into account
the discreteness of the number of individuals: This ensures that
$h_g(x)$ cannot take values less than $1/N$. The evolution equation in
the case of model~B reads
\cite{BrunetDerrida.97}
\begin{equation}
h_{g+1}(x)=\begin{cases}
\displaystyle\min\left(
1,\int d\epsilon\,\psi(\epsilon)h_g(x-\epsilon)
\right)&\text{if that number is larger than $1/N$}\\
0&\text{otherwise}.
\end{cases}
\label{cutoff}
\end{equation}
In the exponential model ($\psi(\epsilon)=e^{-\epsilon}$), it is easy
to see that the solution to (\ref{cutoff}) is
\begin{equation}
h_g(x)=\begin{cases}
1 & \text{for $x<Y_g$},\\
e^{-(x-Y_g)} & \text{for $Y_g<x<Y_g+\ln N$},\\
0 & \text{for $x>Y_g+\ln N$},
\end{cases}
\label{hg_cutoff}
\end{equation}
where the parameter $Y_g$ can be used as the definition of the position
of the front. Substituting (\ref{hg_cutoff}) into (\ref{cutoff}), one
obtains the velocity
\begin{equation}
v_\text{cutoff}^\text{exp}=Y_{g+1}-Y_g=\ln(\ln N+1)\simeq\ln\ln N,
\label{vcutoffexp}
\end{equation}
which does agree, to leading order, with the exact
expression~(\ref{resleadingexpo}).

For a front in the Fisher-KPP class, the cutoff theory 
can also be worked out \cite{BrunetDerrida.97}. One obtains
\begin{equation}
h_g(x)\propto L_0 \sin\left(\pi{x-Y_g\over
L_0}\right)e^{-\gamma_0(x-Y_g)}\qquad\text{and}\qquad
v_\text{cutoff}^\text{F-KPP}=Y_{g+1}-Y_g\simeq
v(\gamma_0)-\frac{\pi ^2
v ^{\prime\prime}(\gamma_0)}{2L_0^2},
\label{cutoffFKPP}
\end{equation}
where $v(\gamma)$ is given by (\ref{vg}), $\gamma_0$ is the value of
$\gamma$ which minimizes $v(\gamma)$, and $L_0=(\ln N)/\gamma_0$ is the
length of the front, from the region where $h_g$ is of order 1 to the
region where it
cancels. The expression of $h_g(x)$ in (\ref{cutoffFKPP}) is
only valid for $h_g(x)\ll1$ and $x-Y_g<L_0$.

By convention, we shall define $\gamma_0=1$ in the exponential case.
Then, both in (\ref{hg_cutoff}) and in (\ref{cutoffFKPP}), the front has
essentially an exponential decay with rate $\gamma_0$ and its length is
$L_0=(\ln N)/\gamma_0$.

So far, (\ref{vcutoffexp}) and (\ref{cutoffFKPP}) have been obtained from
a purely deterministic calculation, where only the discreteness of
$h_g(x)$ has been taken into account. Stochasticity may be put back in
the picture for the generic (Fisher-KPP) case in the following way
\cite{BDMM.06}:

From time to time, a rare fluctuation sends a few individuals ahead of
the front at a distance $\delta$ from its tip. This occurs during the
time interval $dt$ with a probability $p(\delta)\ d\delta\,dt$ where
$p(\delta)$ was assumed \cite{BDMM.06} to be
\begin{equation}
p(\delta)=C_1 e^{-\gamma_0\delta}
\label{pdelta}
\end{equation}
for $\delta$ large enough. $C_1$ is a given constant. 

These individuals then multiply and build up their own front in an
essentially deterministic way. After about $L_0^2$ generations, the
descendants of these individuals have mixed up with the individuals that
stem from the rest of the front. The effect of this rare fluctuation is
therefore to pull ahead the front by a quantity~$R(\delta)$ which,
in the generic (Fisher-KPP) case, is given \cite{BDMM.06} by
\begin{equation}
R(\delta)=\frac{1}{\gamma_0}\ln\left(1+C_2 
\frac{e^{\gamma_0\delta}}{L_0^\alpha}\right),
\label{Rdelta}
\end{equation}
where $C_2$ is another constant and $\alpha=3$.
Finally, in \cite{BDMM.06} it was argued that
\begin{equation}
C_1C_2=\pi^2\gamma_0 v^{\prime\prime}(\gamma_0).
\label{C1C2}
\end{equation}

As we shall show in the next section, the same picture
applies to the exponential model with
some slight modifications: in (\ref{Rdelta}), one
needs to take $\alpha=1$ instead of $\alpha=3$, everywhere $\gamma_0$
must be replaced by 1, one should replace~(\ref{C1C2}) by $C_1=C_2=1$
and the relaxation time of a fluctuation by 1 instead of $L_0^2$.

With these ingredients, it is not difficult to write the generating
function of the position $Y_g$ of the front:
\begin{equation}
\big\langle e^{-\beta Y_g}\big\rangle\sim e^{g
G(\beta)}\quad\text{where}\quad
G(\beta)=-\beta v_\text{cutoff}+\int d\delta\ p(\delta)
\left(e^{-\beta R(\delta)}-1\right).
\label{gendelta}
\end{equation}
The first term in $G(\beta)$ is due to the deterministic motion, while
the integral represents  the effect of the forward rare fluctuations. In
the case of the exponential model, this expression leads
to~(\ref{resleadingexpo}), up to terms of order $1/\ln N$ for the
velocity and of order $\ln\ln N/\ln^2N$ for the other cumulants. In the
generic Fisher-KPP case, the average front velocity, diffusion constant
and higher order cumulants are found from (\ref{gendelta}) to be
\cite{BDMM.06}
\begin{equation}
\begin{split}
v_{N}&=v(\gamma_0)-\frac{\pi ^2\gamma_0 ^2 
v ^{\prime\prime}(\gamma_0)}{2\ln ^2 N}
+\gamma_0 ^2 v ^{\prime\prime}(\gamma_0)\pi ^2
\frac{3\ln\ln N}{\ln ^3 N}+\cdots
=v(\gamma_0)-\frac{\pi ^2\gamma_0 ^2
v ^{\prime\prime}(\gamma_0)}{2(\ln N+3\ln\ln N)^2}+\cdots
,\\
D_N&=\gamma_0 v ^{\prime\prime}(\gamma_0)
\frac{\pi ^4}{3\ln ^3 N}+\cdots,\\
\frac{\left\langle(Y_g-Y_0)^n\right\rangle_c}{g} & =
\gamma_0 ^{3-n} v ^{\prime\prime}(\gamma_0)
\frac{\pi ^2 n!\zeta(n)}{\ln ^3 N}+\cdots\qquad\text{for $n\ge2$}.
\end{split}
\label{resultcumulants}
\end{equation}

One important aspect of (\ref{Rdelta}) is that when $\delta$ is of
order $(\alpha\ln L_0)/\gamma_0$, the
front is shifted by one additional unit in position due to this
fluctuation. This means that a large fraction of the population
is replaced by the descendants of the individuals produced by
this fluctuation. Thus, when one considers a given number of
individuals at generation $g$, the most probable is that their most
recent
common ancestor belongs to one of these fluctuations that triggered
shifts of order 1 in the position of the front in the past generations.
According to~(\ref{pdelta}), such events occur once every $\Delta g \sim
L_0^\alpha$ generations. $\Delta g$ is likely to give the order of
magnitude of the average coalescence times. In section~\ref{gentree}, we
shall build on this to obtain the statistics of the
genealogical trees and the coalescence times 
in the generic Fisher-KPP case. But first, we show that this
phenomenological picture is consistent with the exact results
(\ref{resleadingexpo}) for the exponential model.

\subsection{Exponential case}
\label{ExpCase}

Since the exponential model can be solved exactly (section~\ref{exact}),
one can test in this case our phenomenological picture of
section~\ref{phenoA}. Let us first show that (\ref{pdelta}) gives the
correct distribution of fluctuations.

In the exponential case at any generation $g$, the front is built
according to (\ref{def_yk}) by drawing $N$ independent exponential random
numbers $y_k$ which represent the positions of the particles relative to
a common origin $x_{N+1}$. There is a probability $(1-e^{-y})^N$ that
none of the $y_k$ are on the right of $y$; therefore the distribution
of the rightmost $y_k$ is
\begin{equation}
\prob(y_\text{rightmost})=N\left(1-e^{-y_\text{rightmost}}\right)^{N-1}
e^{-y_\text{rightmost}}
 \simeq\exp\left[{-(y_\text{rightmost}-\ln N)-e^{-(y_\text{rightmost}-\ln N)}} \right].
\end{equation}
$y_\text{rightmost}$ is the distance between the rightmost particle
and the $N+1$-st rightmost particle (before selection). We define
the length $l$ of the front as $l=y_\text{rightmost}$. (A more natural
definition could have been the distance between the
rightmost and the leftmost particles, which is obtained by replacing $N$ by
$N-1$ in the previous equation. For large $N$,
this difference between these two definitions is negligible.)
The average length of the front is therefore
$\langle l\rangle\simeq\ln
N+\gamma_E$ with fluctuations of order 1 given by a Gumbel
distribution, and the probability to observe a large fluctuation where 
$l=\ln N +\delta$ with $\delta\gg1$ is given by
\begin{equation}
p(\delta)\simeq \exp\left[{-\delta-e^{-\delta}} \right]
\simeq \exp\left[-\delta\right],
\end{equation}
which is the same as (\ref{pdelta}).

We now wish to know the effect of such
a fluctuation on the
position of the front. As the shape of the front is decorrelated between
two successive
generations, the relaxation time of a fluctuation is 1 and it is
sufficient to compute $\Delta X_g$ given the value of $\delta$ at
generation $g$. Given the value of $l=y_\text{rightmost}$, the distribution
(\ref{def_yk}) of the $N-1$ other $y_k$ become
\begin{equation}
\prob(y_k)={e^{-y_k}\over1-e^{-l}}\qquad\text{for
$0<y_k<l$.}
\label{py_kgiven}
\end{equation}
As in (\ref{G}), we introduce the generating function of the displacement
$\Delta X_g$ given the value of $l$:
\begin{equation}
\begin{aligned}
\left\langle e^{-\beta\Delta X_g}\middle|l\right\rangle
&=\int dz\, \prob(z) e ^{-\beta z}
\int dy_{1}\,\prob(y_1)\cdots
\int dy_{N-1}\,\prob(y_{N-1})
\left(e^{ y_1}+\cdots+e^{ y_{N-1}}+e^l\right)^{-\beta},\\
&={\Gamma(N+1+\beta)\over\Gamma(N+1)}\,
{1\over(1-e^{-l})^{N-1}}
\int_0^l dy_{1}\,e^{-y_1}\cdots
\int_0^l dy_{N-1}\,e^{-y_{N-1}}
\left(e^{ y_1}+\cdots+e^{ y_{N-1}}+e^l\right)^{-\beta},
\end{aligned}
\end{equation}
where (\ref{intyN}) and (\ref{py_kgiven}) were used.
By using the same representation (\ref{Gammarep}) that led to (\ref{Gint}), one gets
\begin{equation}
\left\langle e^{-\beta\Delta X_g}\middle|l\right\rangle
={\Gamma(N+1+\beta)\over\Gamma(N+1)\Gamma(\beta)}
\int_0^{\infty}d\lambda\,\lambda^{\beta-1}\left[{1\over1-e^{-l}}\int_0^l
dy\,e^{-y-\lambda e^y}\right]^{N-1}e^{-\lambda e^l}.
\end{equation}
which, in terms of $I_0(\lambda)$ defined in (\ref{defI0}), is the same as
\begin{equation}
\left\langle e^{-\beta\Delta X_g}\middle|l\right\rangle
={\Gamma(N+1+\beta)\over\Gamma(N+1)\Gamma(\beta)}
\int_0^{\infty}d\lambda\,\lambda^{\beta-1}\left[I_0(\lambda)-e^{-l}I_0(\lambda
e^{l})\over1-e^{-l}\right]^{N-1}e^{-\lambda e^l},
\label{DeltaXcond}
\end{equation}
where, using (\ref{expansionI0}), 
\begin{equation}
{I_0(\lambda)-e^{-l}I_0(\lambda
e^{l})\over1-e^{-l}}
=1-\lambda{l\over1-e^{-l}}+\sum_{k=0}^{+\infty}
\frac{(-1)^k}{(k+1)(k+2)!}\lambda^{k+2}{e^{l(k+1)}-1\over1-e^{-l}}.
\label{I0diff}
\end{equation}

Expressions (\ref{DeltaXcond}) and (\ref{I0diff}) are valid for any value
of $l$. We now consider a large fluctuation $l=\ln N+\delta$ with
$1\ll\delta\lesssim\ln\ln N$.
As for (\ref{Gint}), the integral is dominated by values of $\lambda$ of
order $1/[N\ln N]$. Making as before
the change of variable $\mu=\lambda N\ln N$, and dropping all the terms
of order $1/N$, one gets
\begin{equation}
\begin{aligned}
\left[I_0(\lambda)-e^{-l}I_0(\lambda
e^{l})\over1-e^{-l}\right]^{N-1}
&\simeq\exp\left[-\mu\left(1+{\delta\over\ln N}\right)
	+\sum_{k=0}^{+\infty}
\frac{(-1)^k}{(k+1)(k+2)!}\left(\mu\over\ln N\right)^{k+2}{e^{\delta(k+1)}}
\right].
\end{aligned}\end{equation}
We are only interested in the leading order in $1/\ln N$. Dropping higher
order terms, one gets, in (\ref{DeltaXcond}),
\begin{equation}
\begin{aligned}
\left\langle e^{-\beta\Delta X_g}\middle|\delta\right\rangle
&\simeq{\Gamma(N+1+\beta)\over\Gamma(N+1)\Gamma(\beta)}
\ {1\over[N\ln N]^\beta}
\int_0^{\infty}d\mu\ \mu^{\beta-1}
\exp\left[-\mu\left(1+{\delta+e^\delta\over\ln N}\right)\right]
\simeq{1\over[\ln N]^\beta}\left(1+{e^\delta\over\ln
N}\right)^{-\beta},
\end{aligned}
\end{equation}
where (\ref{stirling}) has been used and where $\delta$ was neglected
compared to $e^\delta$.

This means that up to the order $1/[\ln N]$ we are considering, $\Delta
X_g$ given $\delta$ is deterministic with
\begin{equation}\begin{aligned}
\Delta X_g(\delta)&\simeq\ln\ln N+\ln\left(1+{e^\delta\over\ln
N}\right)\simeq v_\text{cutoff}+R(\delta),
\end{aligned}
\end{equation}
where we used (\ref{vcutoffexp}) and (\ref{Rdelta}) with
$C_2=\alpha=\gamma_0=1$.

The phenomenological picture we developed for the generic case is
therefore justified for the exponential case: each rare fluctuation of
size $\delta$ in the length of the front leads to a shift $R(\delta)$,
given by (\ref{Rdelta}), for the position of the front.

\subsection{Genealogical trees}
\label{gentree}

With the above scenario, one can also build a simplified picture for the
evolution of a population. We assume that, at each generation, there is
with a small probability a fluctuation of amplitude $f$ produced by an
individual ahead of the front. The long term effect of this fluctuation
is that a fraction $f$ of the population is replaced by the
descendants of this individual.

\begin{figure}[!ht]
\centering
\includegraphics[width=10cm]{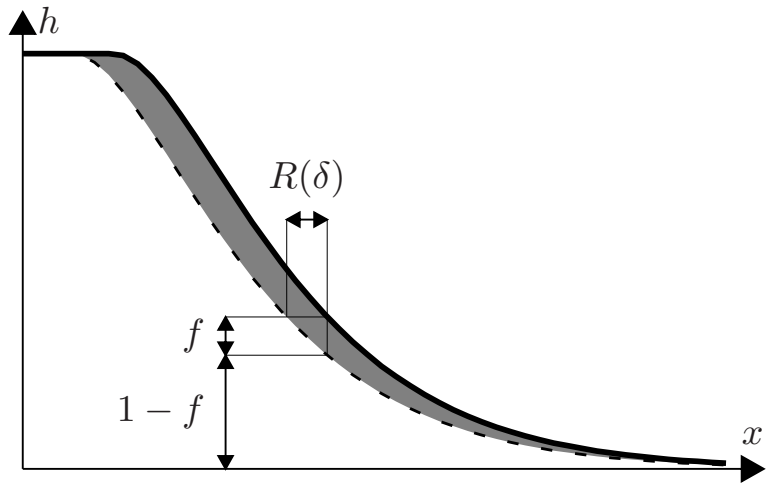}
\caption{Effect of a fluctuation of a front. The dashed line is the front
(\ref{udet}) in the absence of a fluctuation. The plain line is the front
(\ref{ufluct}) if a rare fluctuation occured. The grey area represents
the contribution to the front from the descendants of the fluctuation.
After the front has relaxed,
they represent a proportion $f$ of the whole population.}
\label{figf}
\end{figure}

One can now relate the probability distribution of $f$ to the
phenomenological picture of front propagation. Starting with a front at
position $Y_{g_0}$ at generation $g_0$, we consider its position
$Y_g$ at
a generation $g>g_0$. If no important fluctuation has occurred, the tail
of the front is given by
\begin{equation}
h_\text{no fluctuation}(x,g)\propto
e^{-\gamma_0\big(x-Y_g^\text{no fluctuation}\big)}\quad\text{with }
Y_g^\text{no fluctuation}=Y_{g_0}+v_\text{cutoff}(g-g_0).
\label{udet}
\end{equation}
(See (\ref{cutoffFKPP}); for simplicity, we neglect the Sine prefactor in
the tail as it is a slow varying factor which, to the leading order, does
not change our final result.)

If instead a fluctuation has occurred, generated by an individual ahead
of the front by a distance $\delta$, then the shape is eventually
described by 
\begin{equation}
h_\text{fluctuation}(x,g)\propto
e^{-\gamma_0\big(x-Y_g^\text{fluctuation}\big)}\quad\text{with }
Y_g^\text{fluctuation}=Y_{g_0}+v_\text{cutoff}(g-g_0)+R(\delta).
\label{ufluct}
\end{equation}
that is, the front is pulled ahead  by $R(\delta)$.
If one assumes that the extra mass in the front with fluctuation
(in grey in figure~\ref{figf}) is
due to the fraction $f$ of descendants originating from the fluctuation,
then one gets $h_\text{no fluctuation}=(1-f)h_\text{fluctuation}$.
The substitution of (\ref{udet}) and (\ref{ufluct}) yields
\begin{equation}
f=1-e^{-\gamma_0 R(\delta)}. 
\end{equation}
This equation defines the mapping between the $f$ and the $\delta$
representations of the phenomenological model.
The probability distribution of $\delta$ in (\ref{pdelta})
and the expression (\ref{Rdelta}) of $R(\delta)$
implies the following
distribution of $f$:
\begin{equation}
\prob(f)=\frac{C_1C_2}{\gamma_0 L_0^\alpha}\frac{1}{f^{2}}.
\label{distf}
\end{equation}
(Note that this expression cannot be valid down to $f=0$ for the
distribution to be normalized. One should therefore consider that
(\ref{distf}) is valid above a certain small threshold $f_\text{min}$.
This threshold has no effect on the correlations calculated below.)

Using (\ref{C1C2}) and $\alpha=3$ in the Fisher-KPP case, and
$C_1=C_2=\gamma_0=\alpha=1$ in the exponential case (see
section~\ref{ExpCase}), one gets
\begin{equation}
\prob(f)=
\begin{cases}
\displaystyle
\frac{1}{\ln N}\frac{1}{f^2} & \text{for the exponential model},
\vspace{5pt}\\
\displaystyle
\frac{\pi^2\gamma_0^3 v^{\prime\prime}(\gamma_0)}{\ln^3 N}
\frac{1}{f^2} & \text{for the generic Fisher-KPP case}.
\end{cases}
\label{probf}
\end{equation}
In this model, $p$ individuals may coalesce if they belong to the
fraction $f$ of individuals that are the descendants of a fluctuation.
The probability of such an event thus reads
\begin{equation}
q_p=\int_0^1 df \prob(f)f^p=\frac{C_1C_2}{\gamma_0 L_0^\alpha}\frac{1}{p-1}
\end{equation}
which, for the exponential model, is identical to the exact 
asymptotic result in (\ref{qp}).

The coalescence probabilities in one generation $r_p(k)$ may be obtained
in a straightforward way in this model. One first chooses the $k-1$
individuals among $p$ that do not have a common ancestor in the previous
generation. The latter have to be part of the fraction $1-f$ of
individuals, while the remaining $p-k+1$ individuals that have their
common ancestor in the previous generation must belong to the fraction
$f$. Thus
\begin{equation}
r_p(k)=
\left(\begin{matrix}p\\k-1\end{matrix}\right)
\int_0^1 df\ \prob(f) f^{p-k+1}(1-f)^{k-1}
=\frac{C_1C_2}{\gamma_0 L_0^{\alpha}}
\frac{p}{(p-k)(p-k+1)},
\label{rpf}
\end{equation}
with the same result as in (\ref{rp}) for the exponential model.\footnote{%
In the language of the transition rates $\lambda_{b,q}$ defined in
\cite{Pitman.99,Schweinsberg.00}, one would write
$\lambda_{b,q}=\int_0^1 df\ p(f) f^q (1-f)^{b-q}\propto \int_0^1 df\
f^{q-2}(1-f)^{b-q}$. It is the $\Lambda$-coalescent with the uniform
measure, \textit{i.e.} the Bolthausen-Sznitman coalescent.} At this point,
the combinatorics to get the coalescence probabilities and average times
are the same as in the exact calculation for the exponential model
in section~\ref{sec:treesexp}. 
So, for the exponential model we recover the results of
section~\ref{sec:treesexp} and for
the generic Fisher-KPP case, we get instead
\begin{equation}
\langle T_2\rangle \simeq \frac{\ln^3 N}
{\pi ^2\gamma_0 ^3 v^{\prime\prime}(\gamma_0)},
\label{T2gen}
\end{equation}
while the ratios $\langle T_i\rangle/\langle T_2\rangle$
are the same (\ref{Tpexplicit}) as for the exponential model, in
agreement with the results of numerical simulations 
of \cite{BDMM2.06} and of section~\ref{sec:numerical} below.
Indeed, the $r_p(k)$'s given in (\ref{rp}) and~(\ref{rpf})
are identical except for an overall constant which cancels out
in the ratios.

We note an interesting relation between the average coalescence time and
the front diffusion constant, valid both in the exponential model and in
the generic Fisher-KPP case:
\begin{equation}
D_N\times\langle T_2\rangle\simeq\frac{\pi^2}{3\gamma_0^2}.
\label{DtimesT}
\end{equation}
We will test numerically this identity in section~\ref{sec:numerical}.

As a side remark,
we note that if $\prob(f)$ of (\ref{probf}) is replaced by
$\text{Cste}\, f^{-a}$ with $a\to3$
(instead of $a=2$ in our selective evolution models), 
then the ratios of the coalescence times 
are identical to
those obtained for evolution models without selection, 
see appendix~\ref{appC}.


\section{Numerical simulations}
\label{sec:numerical}

\subsection{Algorithms}

In order to measure the velocity and diffusion constant of our models, it
is sufficient to follow the evolution of the positions of the
individuals. In the case of
model~$A$, at each generation,
one first draws at random the $k$ offspring of each individual 
 and then one keeps the $N$ rightmost offspring as the new population.
This can be done in a computer time linear in $N$.
For model~$B$, one can start by drawing at random the two rightmost
offspring of each individual. If $Z$ is the position of the $N$-th
rightmost offspring out of this first set of $2N$, then one
draws for each individual all its remaining offspring which are larger than
$Z$. Then, taking the $N$ rightmost individuals among those drawn
gives the new population.

We measured the diffusion constants $D_N$ as in \cite{BrunetDerrida.01},
using
$D_N=\big\langle (X_{g_0+g}-X_{g_0}-vg)^2\big\rangle/g$ for a large $g$.
(This
expression is in principle only valid in the $g\to\infty$ limit.) In practice,
we have to choose an appropriate value of $g$ and average over many runs.
For each value of $N$, we measured the diffusion constant twice, once
with $g\approx2\ln^3N$ and once with $g\approx10\ln^3N$, and we have
plotted both values with the same symbol. The fact that one cannot distinguish
the two sets of data indicates that the values of $g$ we took are large
enough and that we accumulated enough statistics.

To measure the statistics of the genealogical trees in the population,
one needs to memorize more information than simply the positions of the
individuals in the current generation. The most na\"ive method would be
to record the whole history of the population, keeping for all
individuals in all generations their positions and parents, and then to
analyze at the end the whole genealogical tree. This is clearly too time and
memory consuming. Instead, we  used the three following algorithms.

The first algorithm consists in working with a matrix $T_g$, the element
$T_g(i,j)$ being the age of the most recent common ancestor of the pair
of individuals $i$ and $j$ at generation $g$. This matrix is simple to
update: if
$j$ and $j'$ are the parents of $i$ and $i'$, then
$T_{g+1}(i,i')=1+T_g(j,j')$ for $i\ne i'$ and
$T_{g+1}(i,i)=0$. By sampling random elements of the matrix at different
generations, one obtains the average value of the
coalescence time between two individuals.
The nice thing is that, due to the ultrametric structure of the tree
(for any $i$, $j$ and $k$, $T_g(i,j)\le\max\big[T_g(i,k),T_g(j,k)\big]$),
no more information is needed to compute the coalescence times of three
or more individuals: the age of the most recent ancestor of $p$
individuals $i_1,\dots,i_p$ is simply given by
$\max\big[T_g(i_1,i_2),T_g(i_1,i_3),\dots,T_g(i_1,i_p)\big]$. This method
is appropriate for values of $N$ up to about $10^3$ as it takes a long
time of order $N^2$ to update the matrix at each generation.

In the second algorithm, instead of working with this matrix $T_g(i,j)$,
we take advantage of the tree structure of the genealogy by recording
only its ``relevant'' nodes: at generation $g$, we say that a node is
``relevant'' if it is an individual of the current generation $g$ or if
it is the first common ancestor of any pair of individuals of the current
generation. Clearly, the ``relevant'' nodes have a tree structure (the
first common ancestor of any two ``relevant'' nodes is a
``relevant'' node), which we record as well.
The leaves of this tree are the current generation, and the root is the
most recent common ancestor of the whole population. This tree
is simple to update:
if, after one timestep, a node has no child, it is removed and its
parent is updated. If a node has only one child, it is removed as well
and its child and parent get directly connected. If the root of the tree
has only one child, it is removed and its child becomes the new root. As
can be seen easily, the tree has at most $2N-1$ nodes and it can be
updated in a time of order $N$. The extraction of the interesting information
from the tree is also very fast: if a node has $p$ children, and these
children are the ancestors of $\alpha_1,\dots,\alpha_p$ individuals of
the current generation, then this node is the most common ancestor of
$\sum_{i\ne j}\alpha_i\alpha_j$ pairs of individuals. More generally, this
node is the most common ancestor of
$\binom{\sum_i\alpha_i}{q}-\sum_i\binom{\alpha_i}{q}$ groups  of $q$
individuals in the current generation. By computing this quantity on each
node of the tree, one obtains the average (or even the distribution) of
all the coalescence times within the current generation in a computer
time of order~$N$. This algorithm turns out to be very fast and we used it
for $N$ up to about $10^6$.

The third algorithm only works for a limited class of models, for which
the positions $x_i(g)$ are integers:
instead of recording the $N$ positions, one only needs to record the
number of individuals at a given site. The typical width of the front
and, therefore, the number of variables to handle, are only of
order $\ln N$.
Let us, then, consider model~B with
$\psi(\epsilon)$ given as a sum of Dirac functions:
$\psi(\epsilon)=\sum_q \phi_q \delta(\epsilon-q)$. This means that, before
selection, an individual at position $x$ has a number of offspring at
position $x+q$ which has a Poisson distribution of average $\phi_q$.
Considering now the whole population, the number of offspring at time
$g+1$ and site
$y$ is also a random Poisson number of average $\sum_{x} n(x,g)
\phi_{y-{x}}$, where $n({x},g)$ is the number of individuals at
site ${x}$ and
time $g$ (compare to (\ref{bigfish})).
To simplify, we consider only cases where
$\phi_q=0$ for $q$ larger than some $q_0$, so that one can easily update
the system from right to left by drawing Poisson numbers and stopping
when the
total number of individuals at time $g+1$ reaches $N$. So far, the method
described allows us to update the positions of the particles, and therefore
to extract the velocity and the diffusion constant, in a time
proportional to $\ln N$ per generation. A similar method has already been used
in \cite{BrunetDerrida.97,BrunetDerrida.01} to simulate populations up to 
$N\simeq10^{100}$.
To extract the coalescence times, one needs to keep more information.
The difficulty resides in the fact that the many individuals at a
given position usually have different ancestors. 
To overcome this difficulty, one can consider the average
coalescence times
$\overline{T}_g(x,x')$ 
of two different individuals at respective positions $x$
and $x'$.
To update that matrix, one starts from the probability that an individual
of generation $g+1$ and position $y$ is the offspring of an individual
who was at position ${x}$:
\begin{equation}
\prob(\text{$y$ comes from ${x}$})=
{n({x},g)\phi_{y-{x}}\over\sum_{x'}n(x',g)\phi_{y-x'}}.
\end{equation}
(Compare to (\ref{Wi_general}).) Then, one obtains that
\begin{equation}
\overline{T}_{g+1}(y,y')=1+\sum_{{x},{x'}} \prob(\text{$y$ comes
from ${x}$}) \prob(\text{$y'$ comes from ${x'}$})
\overline{T}_g({x},{x'})\left(1-{\delta_{{x}}^{x'}\over
n({x},g)}\right).
\label{updT2}
\end{equation}
(The term in parenthesis is the probability that individuals at positions
$y$ and $y'$ come from
two different parents given the parents' positions $x$ and $x'$.)
Then, the average
coalescence time of two individuals in the population is simply given by
\begin{equation}
{1\over N(N-1)}\sum_{x,x'} \overline{T}_{g}(x,x') n(x,g) n(x',g)
\left(1-{\delta_{x}^{x'}\over n(x,g)}\right).
\end{equation}
Therefore, by storing a matrix of size $\ln^2 N$ which can be updated in
a time $\ln^4 N$, one can obtain the average coalescence time of two
individuals. An interesting observation is that this algorithm simulates
one possible realization of the positions of the particles; however, the
quantity $\overline{T}_{g}(x,y)$ is actually an average over all the
possible genealogical trees in the population given that realization of
the positions over time of the particles. A complexity in time of
order $\ln^4 N$ allows already to simulate rather large systems.
However, a further optimization is possible in the special case where
$\phi_q$ is constant for $q\le q_0$. For that specific model,
additional simplifications occur (one can write a recursion on the matrix
elements) and the matrix $\overline{T}_{g}(x,x')$ can be
updated in a time of only $\ln^2 N$. This allows one to study systems of
size $N$ up to about $10^{50}$ in a few weeks time on standard desktop
computers. There is, unfortunately, not enough information in the
matrix $T_g(x,x')$ to extract the average coalescence time
of three (or more) individuals: to that purpose, one  needs to simulate a
tensor with three (or more) indices which can be updated with rules very
similar to (\ref{updT2}). Because of this extra complexity, we only measured
the average coalescence time
of three individuals for values of $N$ up to $10^{20}$. 

\subsection{Results}\label{sec:numresults}

\begin{figure}[!ht]
\centering
\includegraphics[width=12cm]{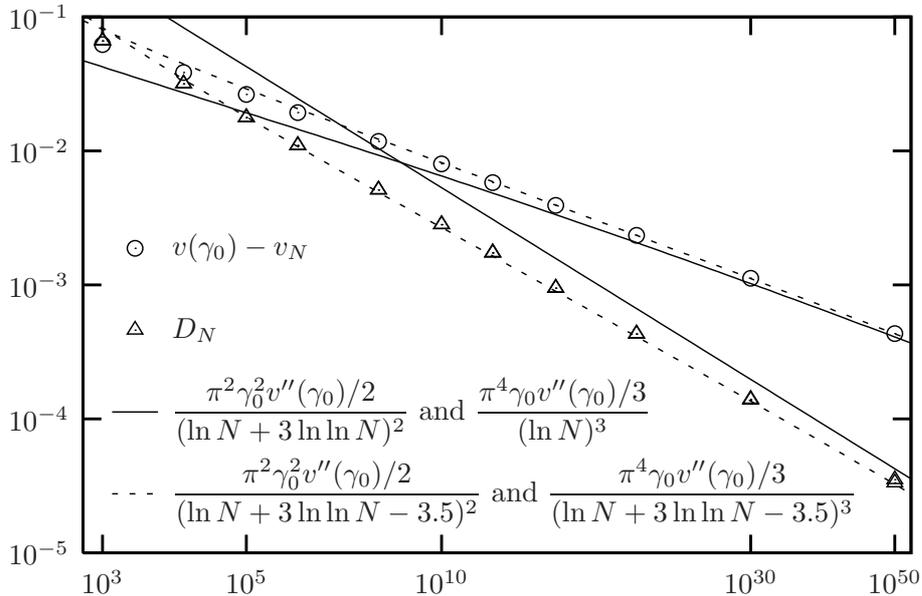}
\caption{Numerical simulations of model~B with
$\psi(\epsilon)={1\over4}\sum_{n\le0}\delta(\epsilon-n)$. The circles are
the correction to the velocity and the triangles the diffusion constant,
as a function of $N$. The plain lines are the predictions
(\ref{resultcumulants}). The dotted lines are the predictions
(\ref{resultcumulants}) with,  for both quantities,
the same subleading terms added in the denominators. (The scale on
the $N$ axis is proportional to $\ln\ln N$.)}
\label{fig:Dv}
\end{figure}

Using this last algorithm, we have simulated model~B for
$\psi(\epsilon)={1\over4}\sum_{n\le0}\delta(\epsilon-n)$ up to
$N=10^{50}$. The velocity and diffusion constants are shown in
figure~\ref{fig:Dv}, compared to the predictions
(\ref{resultcumulants}) in plain lines. There is still a small visible
difference between numerics and theory, but this difference gets smaller
as $N$ increases.  In order to
obtain a better fit, we have included subleading corrections by
changing the denominator $(\ln N+3\ln\ln N)^2$ for the velocity in
(\ref{resultcumulants}) into $(\ln N+3\ln\ln N-3.5)^2$. Similarly, we
changed the denominator $(\ln N)^3$ for the diffusion constant in
(\ref{resultcumulants}) into $(\ln N+3\ln\ln N-3.5)^3$. With these
subleading terms (in dotted lines on the figure), the fit is almost
perfect over more than 40 orders of magnitude.

We have no theory to justify these extra subleading terms, but we simply
notice that it is possible to fit both the correction to the velocity and
the diffusion constant using the same subleading terms in the
denominators of their respective expressions.

\begin{figure}[!ht]
\centering
\includegraphics[width=12cm]{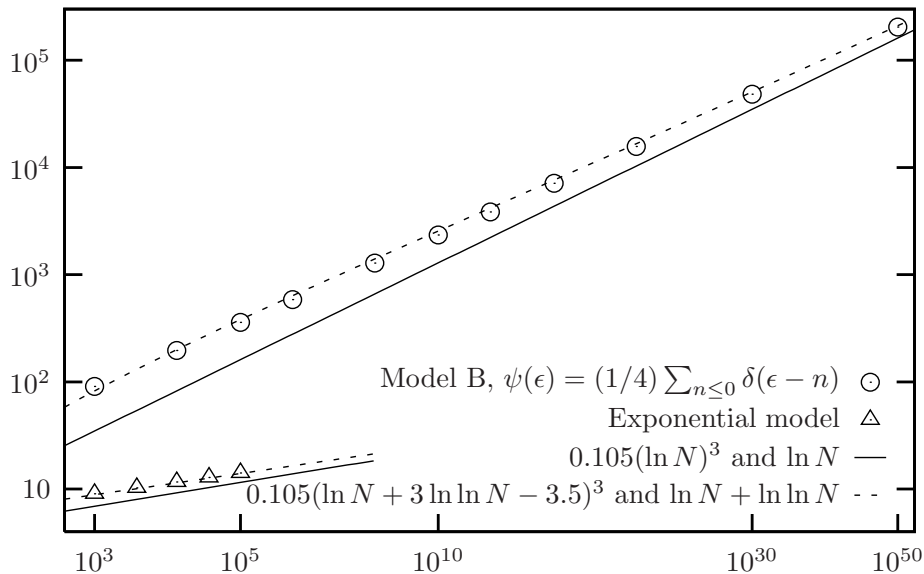}
\caption{Numerical simulations of $\langle T_2\rangle$ for model~B with
$\psi(\epsilon)={1\over4}\sum_{n\le0}\delta(\epsilon-n)$ (circles) and
for the exponential model (triangles). The plain lines are the
predictions (\ref{Tpexplicit})
for $\langle T_2 \rangle$ and (\ref{T2gen}), while the dotted lines
are the same predictions with some subleading term: for the generic case,
we used subleading terms suggested by (\ref{DtimesT}) and the fit of
figure~\ref{fig:Dv}, and for the exponential model the exact results
(\ref{Tpappendix}).}
\label{fig:T2}
\end{figure}
For the same model, $\langle T_2\rangle$ is shown on
figure~\ref{fig:T2} (using circles), compared to the prediction
(\ref{T2gen}) in plain lines. As for the velocity and diffusion constant,
there is still a small visible difference and we obtain a better fit if
we include subleading terms (in dotted lines): guided by (\ref{DtimesT})
and the fit used for the diffusion constant in figure~\ref{fig:Dv}, we
changed the numerator of (\ref{T2gen}) from $(\ln N)^3$ into $(\ln
N+3\ln\ln N-3.5)^3$. On the same figure, $\langle T_2\rangle$ for
the exponential model is shown (using triangles), compared with the exact
prediction
(\ref{Tpexplicit}) $\langle T_2\rangle\simeq\ln N$. Here again, the fit
is improved by including the subleading corrections (\ref{Tpappendix})
$\langle T_2\rangle\simeq\ln N+\ln\ln N$ obtained in appendix~\ref{appA}.

Figure~\ref{fig:ratio} combines data from figures~\ref{fig:Dv} and~\ref{fig:T2}. The
triangles are the ratio of the diffusion constant and of the correction
to the velocity to the power $3/2$. For large~$N$, this should converge
to a constant which we can compute from (\ref{resultcumulants}). The
circles are the product of the diffusion constant and of the coalescence
time~$\langle T_2\rangle$, which we expect to converge to the value given
in (\ref{DtimesT}). The horizontal lines on the figure represent both
predictions.

\begin{figure}[!ht]
\centering
\includegraphics[width=12cm]{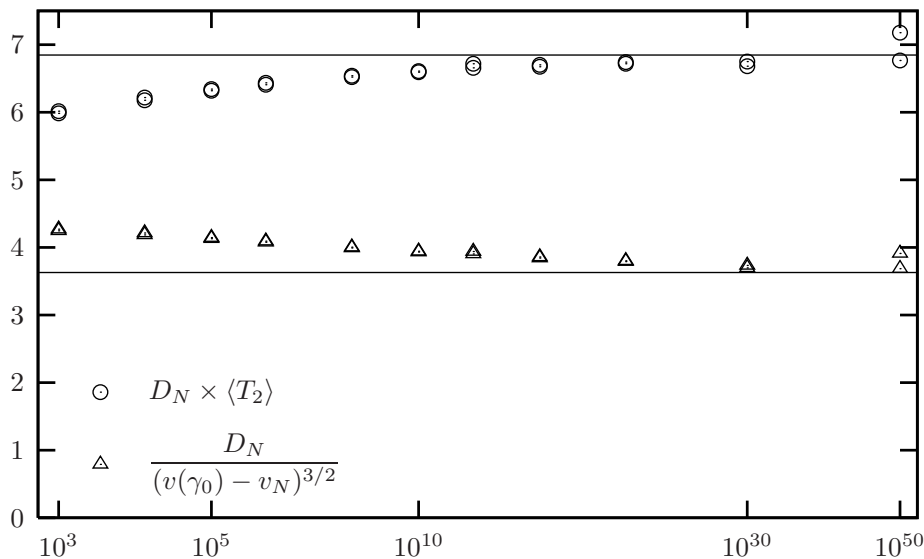}
\caption{Numerical simulations of model~B with
$\psi(\epsilon)={1\over4}\sum_{n\le0}\delta(\epsilon-n)$. The circles
represent the product $D_N\times\langle T_2\rangle$ compared to the
prediction (\ref{DtimesT}). The triangles are the ratio of the diffusion
constant and the correction to the velocity to the power 3/2, compared to
$\pi\sqrt{8/v''(\gamma_0)}/(3\gamma_0^2)$, which is the prediction
obtained from (\ref{resultcumulants}).}
\label{fig:ratio}
\end{figure}

\begin{figure}[!ht]
\centering
\includegraphics[width=12cm]{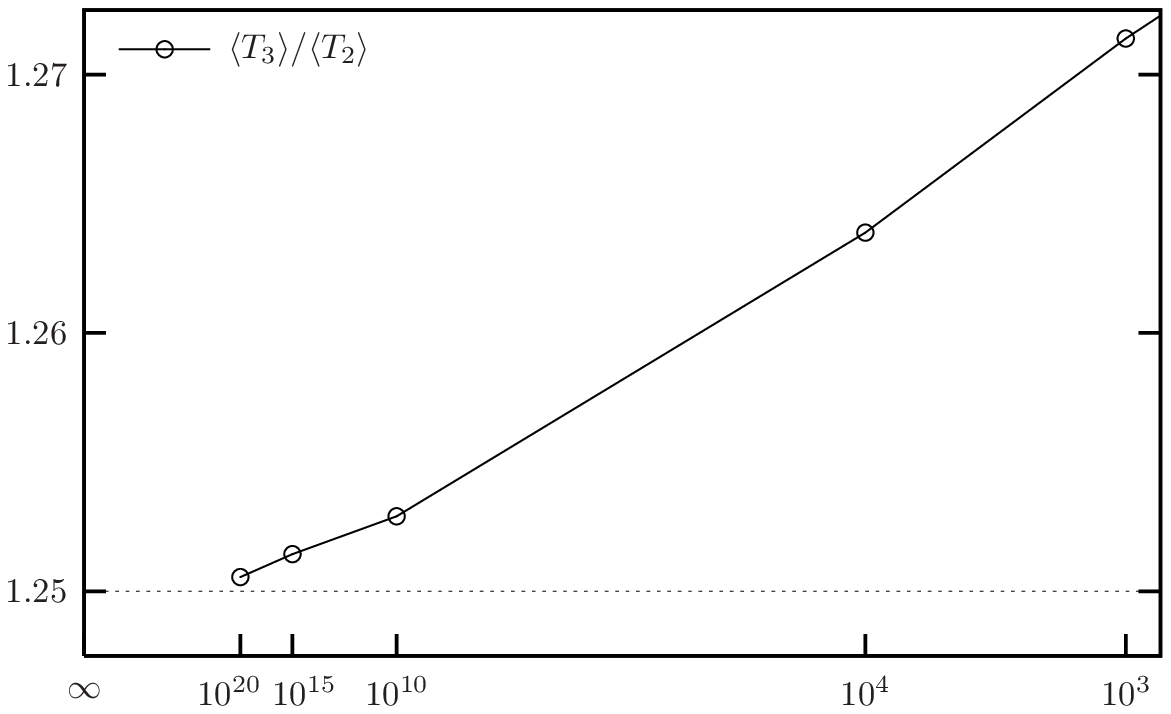}
\caption{Numerical simulations of model~B with
$\psi(\epsilon)={1\over4}\sum_{n\le0}\delta(\epsilon-n)$. The circles
represent the ratio $\langle
T_3\rangle/\langle T_2\rangle$ as a function of $N$, compared to
the result $5/4$ suggested by the phenomenological theory of
section~\ref{gentree}. (The scale on the $N$ axis is proportional to
$1/\ln N$.)}
\label{fig:T3}
\end{figure}

Finally, figure~\ref{fig:T3} shows the ratio $\langle
T_3\rangle/\langle T_2\rangle$ as a function of $N$ up to $N=10^{20}$.
The ratio is very close to 1.25 for large $N$, which is the prediction
of the phenomenological theory of section~\ref{gentree} (see
also (\ref{Tpexplicit})).

\section{Conclusion}

In the present work, we have solved exactly a simple model of evolution
with selection, the exponential model of section~\ref{exact}. For this
model, we have calculated the velocity and the diffusion constant
(\ref{resleadingexpo}) of the parameter
representing the adequacy of the population to its environment,
as well as the coalescence times which characterize the genealogy. We
have shown that the statistical properties of the genealogical trees are
identical to those trees which appear in the Parisi mean field theory of
spin glasses \cite{MezardPSTV.84,MezardParisiVirasoro.87}. They therefore
follow the Bolthausen Sznitman statistics
\cite{Ruelle.87,BolthausenSznitman.98}, in contrast to the case of
evolution without selection which obeys the statistics of the Kingman
coalescent.

The reason why the exponential model is exactly soluble is that, going
from one generation to the next, the only relevant information on the
position of the individuals is contained in one single variable $X_g$
defined in (\ref{defXg}). The exponential model belongs to a larger class
of models parametrized by a single function $\rho$ (for model~A) or
$\psi$ (for model~B). We have not been able to solve the generic case
and, unfortunately, the exponential model is special: while the generic
case can be described by a Fisher-KPP front, with a velocity which
converges when $N\to\infty$, the velocity of the front associated to the
exponential model diverges when $N\to\infty$. We have however
constructed a phenomenological picture of front propagation which can be
used both for the exponential model and for the generic Fisher-KPP case,
and which also provides predictions for the genealogy. Within this
picture, we have that the average coalescence times scale like $\ln^3 N$
with the size $N$ of the population for the generic Fisher-KPP case
(while it grows like $\ln N$ for the exponential model), and that the
structure of the trees is  the same as in the Parisi
mean-field theory of spin glasses.

Proving the validity of the phenomenological picture for generic models
is an interesting open question for future research. Understanding more
deeply why our models of selective evolution are related to spin
glasses would also deserve some efforts. Lastly, it would be interesting
to study genealogies in other models of selective evolution
\cite{Kessler.97} to test the robustness of our results.


\section*{Acknowledgments}
This work was partially supported by the US Department of Energy.


\appendix

\section{Exact results for the exponential model including subleading orders}
\label{appA}

In this appendix, we obtain higher orders in the large $\ln N$ expansion,
for the statistics of the position of the front and for the coalescence
probabilities in the exponential model.

\subsection{Front position statistics}

The exact expression for the cumulants of the front velocity was given in
(\ref{Gint}) in terms of the function $I_0$ defined in (\ref{defI0}).
Discarding all the terms of order $1/N$ or smaller, one can use directly
the expression (\ref{I0^N}) of $[I_0(\lambda)]^N$ as a function of the
rescaled variable $\mu$ in~(\ref{Gint}). Keeping terms up to the order
$1/\ln^2N$, one gets, using also (\ref{stirling}),
\begin{equation}
e^{G(\beta)}={1\over\ln^\beta N}\,{1\over\Gamma(\beta)}
\int_0^{\infty}d\mu\,\mu^{\beta-1}e^{-\mu}
\left(1+\mu{\ln\mu-\ln\ln N+\gamma_E-1\over\ln N}+
{1\over2}\left[\mu{\ln\mu-\ln\ln N+\gamma_E-1\over\ln
N}\right]^2+\cdots\right).
\end{equation}
The integrals of each term can be computed using (\ref{derGamma}).
One gets
\begin{equation}
e^{G(\beta)}={1\over\ln^\beta N}\left[1+{\beta\over\ln
N}\left({\Gamma'(\beta+1)\over\Gamma(\beta+1)}
-l\right)+
{\beta(\beta+1)\over2\ln^2N}\left(
	{\Gamma''(\beta+2)\over\Gamma(\beta+2)}
	-2{\Gamma'(\beta+2)\over\Gamma(\beta+2)}l
	+l^2
\right)
+ \cdots\right]
\label{expGbappA}
\end{equation}
with $l=\ln\ln N-\gamma_E+1$.
Taking the logarithm of (\ref{expGbappA}), one obtains $G(\beta)$. By
expanding in powers of $\beta$ and comparing with (\ref{defcum}), one
gets the cumulants of the position of the front. We give the velocity and
diffusion constant:
\begin{equation}\begin{aligned}
v_N&=\ln\ln N+\frac{\ln\ln N+1}{\ln N}
-{(\ln\ln N)^2-1+{\pi^2\over6}\over2\ln^2N}+\cdots,\\
D_N&=\frac{\pi^2}{3}\frac{1}{\ln N}
-{1\over\ln^2N}\left({\pi^2\over3}\ln\ln N-{\pi^2\over6}+2\zeta(3)\right)
+\cdots
\end{aligned}
\end{equation}
Note that the first correction to the leading term can be in both cases
obtained by replacing in the leading term $\ln N$ by $\ln N+\ln\ln N$:
$v_N  \simeq\ln(\ln N+\ln\ln N)$ and $D_N\simeq(\pi^2/3)/(\ln N+\ln\ln
N).$ This is reminiscent of the observation in figure~\ref{fig:Dv} that,
in the generic case, the fit was better by replacing the $\ln N$
by $\ln N+3\ln\ln N$ in the theoretical
prediction for the diffusion constant.

\subsection{Tree statistics}

To get subleading orders for the statistics of the tree in the
exponential case, one needs to generalize the discussion in
section~\ref{sec:treesexp} where we derived
the leading term in the large $\ln N$ expansion.
The central quantity is still the probability $r_p(k)$
that $p$ individuals at generation $g+1$ have exactly $k$ ancestors in
the previous generation.
But while at leading order it was enough to consider one coalescence at
each step, one needs to take into account
up to $n$ simultaneous coalescences when one wishes to
keep terms of arbitrary order  $1/\ln^n N$.

One has to assign an ancestor at generation $g$ to each individual 
at generation $g+1$.
We start from the probability $W_i(x)$ given in
(\ref{Wi_general}) that the parent of an individual at position $x$ and
generation $g+1$ was the $i$-th individual of generation~$g$. 
In the exponential model, $W_i(x)$ does not depend on
$x$ (see (\ref{defWi})).
We consider $p$ individuals of generation $g+1$ and we note $p_i$ the
number of these individuals that are descendants of the $i$-th individual
of generation $g$. The probability distribution of the $p_i$ is
\begin{equation}
\prob(p_1,\dots,p_N)={p!\over p_1!\cdots p_N!}\delta_{p_1+\cdots+p_N}^p
W_1^{p_1}\cdots W_N^{p_N}.
\end{equation}
One now averages
over the positions of individuals at generation $g$, and
$r_p(k)$ is simply the probability that there are exactly $k$ non zero
$p_i$'s. After relabeling the individuals at generation $g$, one gets
\begin{equation}
r_p(k) = \binom{N}{k}\sum_{p_1\ge1,\cdots,p_k\ge1}{p!\over p_1!\cdots
p_k!}\delta_{p_1+\cdots+p_k}^p\langle W_1^{p_1}\cdots W_k^{p_k} \rangle.
\end{equation}
It is actually convenient to call $n$ the number of $p_i$ that are
strictly larger than~1 and to write $r_p(k)$ as a sum over~$n$: after
another relabeling,
\begin{equation}
r_p(k)=
\left(\begin{matrix}N\\k\end{matrix}\right)
\sum_{n\ge0}
 \left(\begin{matrix}k\\n\end{matrix}\right)
\sum_{p_1\ge 2,\cdots,p_n\ge 2}
\frac{p!}{p_1!\cdots p_n!}
\delta^{p-k+n}_{p_1+\cdots+p_n}
\left\langle
W_1^{p_1}\cdots W_n ^{p_n}
W_{n+1}\cdots W_k
\right\rangle.
\label{rpknext}
\end{equation}
Indeed, as we shall see, each term in the sum over $n$ gives a
contribution of order $1/\ln^n N$ in the final result. The averaged term
can be expressed using the probability $W_i$ given in (\ref{defWi}):
\begin{equation}
J_{p,k,n}^{p_1,\dots, p_n}=
\langle W_1^{p_1}\cdots W_n^{p_n}W_{n+1}\cdots W_k\rangle
=\int_0^\infty dy_1\ e^{-y_1}\cdots
 \int_0^\infty dy_N\ e^{-y_N}
\frac{e^{p_1 y_1+\cdots+p_n y_n+y_{n+1}+\cdots+y_k}}
{\left(e^{y_1}+\cdots+e^{y_N}\right)^p}.
\end{equation}

The technique to evaluate the integrals involved here
is essentially the same as in section.~\ref{exact}.
We first use the standard representation~(\ref{Gammarep})
for the denominator in the integrand. Then the integral over $y_i$ may be
expressed with the help of the functions $I_p(\lambda)$ defined in
(\ref{defIp}):
\begin{equation}
J_{p,k,n}^{p_1,\dots, p_n}=
\frac{1}{(p-1)!}\int_0^{+\infty} d\lambda\,
\lambda^{p-1}I_{p_1}(\lambda)\cdots I_{p_n}(\lambda)
I_1(\lambda)^{k-n}I_0(\lambda)^{N-k}.
\label{Jp1}
\end{equation}
As before, for large $N$, the term $I_0(\lambda)^N$ make the integral
(\ref{Jp1}) dominated by values of $\lambda$ of order $1/[N\ln N]$.
It is sufficient to use the leading order (\ref{Ipsmall}) for the
$I_p(\lambda)$ as next orders in $\lambda$ would generate terms
of order $1/N$, which we discard throughout.
Making the change of variables  $\mu=\lambda N\ln N$ (see
(\ref{mulambda})), and
using the fact that $p_1+\cdots+p_n=p-k+n$, one
gets for the integrand of (\ref{Jp1})
\begin{equation}
\lambda^{p-1}I_{p_1}\cdots I_{p_n}
I_1^{k-n}I_0^{N-k}
\simeq {(p_1-2)!\cdots(p_n-2)!\over N^{k-1}\ln^{n-1} N}\mu^{k-1} e^{-\mu}
\left[1+{(k-n-\mu)(\ln\ln N-\ln\mu-\gamma_E)-\mu\over\ln N}+\cdots\right]
\end{equation}
(\ref{Jp1}) can then be evaluated using (\ref{derGamma}). One gets
\begin{equation}
J_{p,k,n}^{p_1,\dots, p_n} =
{(p_1-2)!\cdots(p_n-2)!\over(p-1)!}\
{(k-1)!\over N^k\ln^n N}
\left[1+{n\left({\Gamma'(k)\over\Gamma(k)}+\gamma_E-\ln\ln
N\right)-(k-1)\over\ln N}+\cdots\right]
\end{equation}
as expected, $j_{p,k,n}$ has an amplitude proportional to  $1/\ln^n N$.
To compute $r_p(k)$ for $k<p$ to order $1/\ln^2N$, one only needs in (\ref{rpknext}) the terms $n=1$ and $n=2$  (the term $n=0$ gives a contribution only for $k=p$):
\begin{equation}
r_p(k)\simeq {N^k\over k!}\left(k {p!\over(p-k+1)!}J_{p,k,1}^{p-k+1}
+{k(k-1)\over2} \sum_{p_1=2}^{p-k}{p!\over p_1!(p-k+2-p_1)!}
J_{p,k,2}^{p_1,p-k+2-p_1}+\cdots\right).
\end{equation}
After some algebra, one gets, for $k<p$,

\begin{equation}
r_p(k)=\frac{p}{(p-k+1)(p-k)}\frac{1}{\ln N}
\left[
1+\frac{1}{\ln N}
\left(
\sum_{n=1}^{k-1}\frac{1}{n}
+\frac{2(k-1)}{p-k+2}\left(
\sum_{n=1}^{p-k-1}\frac{1}{n}-\frac32
\right)-\ln\ln N
\right)+\cdots
\right].
\end{equation}
(we used, among other things,
$\Gamma'(k)/\Gamma(k)+\gamma_E=1+{1\over2}+\cdots+{1\over k-1}$.)

We can now compute the $\langle T_k\rangle$. From the recurrence
\begin{equation}
\langle T_p\rangle=1+\sum_{k=1}^p r_p(k)\langle T_k\rangle,
\end{equation}
we get, using $\sum_k r_p(k)=1$ and $\langle T_1\rangle=0$,
\begin{equation}
\langle T_p\rangle=
\frac{1+\sum_{k=2}^{p-1}r_p(k)\langle T_k\rangle}
{\sum_{k=1}^{p-1} r_p(k)}.
\end{equation}

For the first values of $p$, we obtain
\begin{equation}
\begin{split}
\langle T_2\rangle&=\ln N+\ln\ln N+o(1)\\
\langle T_3\rangle&=\frac54(\ln N+\ln\ln N)+o(1)\\
\langle T_4\rangle&={25\over18}(\ln N+\ln\ln N)-{1\over 54}+o(1).
\end{split}
\label{Tpappendix}
\end{equation}


\section{The Parisi Broken Replica symmetry}
\label{appB}

The replica trick is a powerful approach to calculate the typical free
energy of a sample in the theory of disordered systems. In the replica
trick, one considers $n$ replicas of the same random sample,  one
averages the product of their  partition functions and at the end of the
calculation one takes the limit $n \to 0$. In some cases, the
$n$-dependence of this averaged product is simple enough for the analytic
continuation $n \to 0$ to be unique leading to the desired free energy.

In the case of mean-field spin-glasses, the situation is more
complicated:  the symmetry between the replicas gets broken as $n$ takes
non-integer values ($n<1$) and remains broken in the limit $n \to 0$. In
this appendix we recall the statistical properties of the trees predicted
by the Parisi theory of the broken replica symmetry
\cite{Parisi.83,Parisi.80,MezardPSTV.84,MezardParisiVirasoro.87}.

One starts with an integer $n=n_0$ number of replicas. These replicas are
grouped into $n_0/n_1$ groups of $n_1$ replicas. Each of these groups of
$n_1$ replicas is decomposed into $n_1/n_2$ groups of $n_2$ replicas and
so on: each group of $n_i$ replicas is formed of $n_i/n_{i+1}$ groups of
$n_{i+1}$ replicas each. When this hierarchy consists of $k$ levels, it
is  characterized by $k+1$ integers
\begin{equation}
n=n_0 > n_1> n_2 > ...   > n_k =1.
\label{ineq}
\end{equation}
At level $i$, there are a total of $n/n_i$ groups of size $n_i$.
Therefore, the probability that $m$ distinct individuals chosen at random
belong to the same group at level $i$ (without specifying whether they
belong or not to the same group at level $i+1$) is
\begin{equation}
Q_m={{n\over n_i}\binom{n_i}{m}\over\binom{n}{m}}=
\frac{n(n_i-1)(n_i-2) \cdots (n_i-m+1)}{n (n-1)\cdots(n-m+1)}
\label{Qn}
\end{equation}
One can also associate a tree to each choice of $m$ replicas: the $m$
replicas are at the bottom of the tree and when two replicas belong to
the same group at level~$i$, but to different groups at level~$i+1$,
their branches merge at level~$i$.

The various possible trees which might occur for three replicas or four
replicas are shown in tables~\ref{trees3} and~\ref{trees4} with their probabilities.
For example for the first tree of table~\ref{trees3}, the probability 
that two branches merge at level $j$ and the remaining branches
merge at level $i$ is 
\begin{equation}
\frac{n(n_i - n_{i+1})(n_j - n_{j+1})}{n (n-1)(n-2)},
\end{equation}
as there are $n$ possible choices for the leftmost replica,
$n_i-n_{i+1}$ choices for the rightmost replica and $n_j-n_{j+1}$ 
choices for the replica at the center of the figure.
The degeneracy factor is simply the number of different ways of permuting
the roles of the replicas at the bottom of the tree.

\begin{table}[!ht]
\centering
\begin{tabular}{m{12mm}m{4.5cm}l}
\includegraphics[width=11mm]{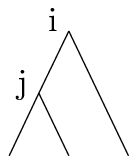}
&$\displaystyle\frac{n(n_i-n_{i+1})(n_j-n_{j+1})}{n(n-1)(n-2)}$&3 cases\\[4ex]
\includegraphics[width=11mm]{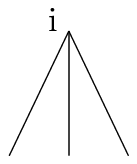}
&$\displaystyle\frac{n(n_i-n_{i+1})(n_i-2n_{i+1})}{n(n-1)(n-2)}$&1 case
\end{tabular}
\caption{All possible trees of three replicas, their probabilities and
degeneracies.}
\label{trees3}
\end{table}

\begin{table}[!ht]
\centering
\begin{tabular}{m{12mm}m{6cm}l}
\includegraphics[width=11mm]{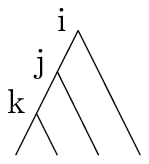}
&$\displaystyle\frac{n(n_i-n_{i+1})(n_j-n_{j+1})(n_k-n_{k+1})}
{n(n-1)(n-2)(n-3)}$&12 cases\\[4ex]
\includegraphics[width=11mm]{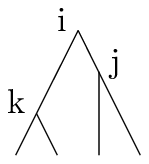}
&$\displaystyle\frac{n(n_i-n_{i+1})(n_j-n_{j+1})(n_k-n_{k+1})}
{n(n-1)(n-2)(n-3)}$&6 cases\\[4ex]
\includegraphics[width=11mm]{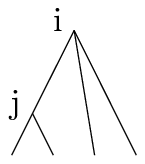}
&$\displaystyle\frac{n(n_i-n_{i+1})(n_i-2n_{i+1})(n_j-n_{j+1})}
{n(n-1)(n-2)(n-3)}$&6 cases\\[4ex]
\includegraphics[width=11mm]{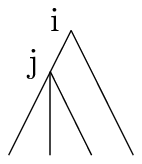}
&$\displaystyle\frac{n(n_i-n_{i+1})(n_j-n_{j+1})(n_j-2n_{j+1})}
{n(n-1)(n-2)(n-3)}$&4 cases\\[4ex]
\includegraphics[width=11mm]{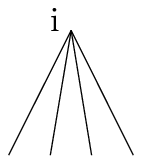}
&$\displaystyle\frac{n(n_i-n_{i+1})(n_i-2n_{i+1})(n_i-3n_{i+1})}
{n(n-1)(n-2)(n-3)}$&1 case
\end{tabular}
\caption{All possible trees of four replicas, their probabilities and
degeneracies.}
\label{trees4}
\end{table}

In the Parisi ansatz, all the calculations are done as if all the $n_i$'s
and all the ratios $n_i / n_{i+1}$ were integers. At the end of the
calculation, however, one takes the limit $n\to 0$ and one reverses the inequality
(\ref{ineq}) into
\begin{equation}
n=n_0 < n_1< n_2 < \cdots   < n_k =1.
\label{ineq1}
\end{equation}
One then  takes a continuous limit ($k \to \infty$)
where $n_i$ becomes a continuous variable $x$
\begin{equation}
n_i=x.
\label{nix}
\end{equation}
In the spin glass theory \cite{MezardPSTV.84,MezardParisiVirasoro.87},
there is an ultrametric distance between pairs of replicas, related to
the overlap $q_{\alpha,\beta}$. (The distance is a decreasing function of
the overlap.) 
This overlap $q_{\alpha,\beta}$ depends on the
level at which the branches of these two replicas merge: this means that
at each level~$i$ of the hierarchy, one associates a value $q_i$ of the
overlap and that $q_{\alpha,\beta}=q_i$ if the two replicas
$\alpha$ and $\beta$ belong to the
same group at level~$i$ and to different groups at level~$i+1$.
($q_i$ is an increasing function of $i$ with $q_0=0$ and $q_k=1$.)
In the limit~$k\to\infty$, when the $n_i$ become a continuous variable
(\ref{nix}), the overlap $q_i$ becomes a increasing
function~$q(x)=q(n_i)=q_i$ with $q(0)=0$ and $q(1)=1$.

The probability that two replicas have an overlap $q_{\alpha,\beta}<q_i$
is
\begin{equation}
\prob(q_{\alpha,\beta}=q_0)+
\prob(q_{\alpha,\beta}=q_1)+\cdots+\prob(q_{\alpha,\beta}=q_{i-1})
=1-Q_2(n_i)={n-n_i\over n-1}.
\end{equation}
Therefore, in the $n\to0$ limit,
the probability $P(q)$ that the overlap $q_{\alpha,\beta}$ between two
replicas $\alpha$ and $\beta$ takes the value $q$ is then given by
\begin{equation}
\int_0^{q(x)}P(q')\ dq' =  \lim_{n \to 0}(1-Q_2) =x
\end{equation}
and this leads to the famous relation \cite{Parisi.83} between the function 
$q(x)$ and the probability distribution of the overlap
\begin{equation}
P(q)= \frac{dx}{dq}.
\end{equation}

In our models, the coalescence time between a pair of individuals in
the population defines, clearly, an ultrametric distance.
In order to see whether the statistics  predicted by the replica approach 
remain valid for the trees of the exponential model
discussed in the present paper, one needs to
relate the overlap $q(x)$ or
the parameter $x$ (which indexes the height of the hierachy)
to the coalescence time $T$ by a function
$T(x)$. It turns out that this 
can be achieved by identifying the probability $e^{-T}$
 that the 
coalescence time  between two individuals is larger than $T$ (see
$R_2(T)$ in (\ref{R_p})) with 
the probability that two replicas belong to different groups
at level $i$.
In other words 
\begin{equation}
e^{-T}=1-Q_2= \frac{n(n-n_i)}{n(n-1)},
\end{equation}
which leads in the $n\to 0$ limit to
\begin{equation}
e^{-T}=x.
\label{identification}
\end{equation}

With this identification, if one assumes that the statistics 
of the trees are given by Parisi's theory, one can compute 
all the statistical properties of the coalescence times of 
trees. For example, by taking the $n\to0$ limit of (\ref{Qn}), one gets
that the probability $Q_m$ that $m$ individuals have a coalescence time
$T_m<T$ is given by
\begin{equation}
Q_m \to \frac{\Gamma(m-x)}{(m-1)! \ \Gamma(1-x) }
\end{equation}
which, by taking the derivative with respect with $T$, gives 
\begin{equation}
\langle (T_m)^p\rangle = \int_0^1 dx  \ T(x)^p \ \frac{d Q_m}{ dx }  =  
 \int_0^\infty dT \  T^p \  \frac{d}{dT}  
\frac{\Gamma(m-e^{-T})}{(m-1)! \ \Gamma(1-e^{-T}) }.
\end{equation}
This coincides 
with the result of the direct calculation~(\ref{Pm}) of the moments of
the $T_m$ and shows that the statistics of the trees in the
exponential model are the same as the ones predicted by the mean field
theory of spin glasses.


\section{The neutral model}
\label{appC}

In this appendix we recall some well known results on the statistical 
properties of the coalescence times in neutral models
\cite{Kingman.82,TavareBGD.97} and derive (\ref{Tneutral}).

We consider a population of fixed size $N$ with non overlaping generations.
Each individual $i$ at a given generation $g$ has $k_i(g)$ offspring
at the next generation. We assume that the $k_i(g)$ are random and
independent, and we call $p_k$ be the probability that $k_i(g)=k$. The
total number $M$ of offspring is therefore given by
\begin{equation}
M= \sum_{i=1}^N k_i.
\end{equation}
To keep the size of the population constant we choose $N$ individuals at random
among these $M$ individuals.

The probability $q_n$ that $n$  individuals have the same parent 
at the previous generation is 
\begin{equation}
q_n =\left\langle\sum_i \binom{k_i}{n}\over\binom{M}{n}\right\rangle
	=\left\langle\frac{\sum_i k_i(k_i-1) \cdots (k_i-n+1)}{ M (M-1)
\cdots (M-n+1)}\right\rangle.
\end{equation}
For a population of large size, if $p_k$ decays
fast enough with $k$ for the moments of $k$ to be finite,
the law of large numbers gives that the denominator is
approximatively equal to $(N\langle k\rangle)^n$ and
\begin{equation}
q_n \simeq  \frac{1}{N^{n-1}\langle k\rangle ^n}
\left\langle
\frac{\Gamma(k+1)}{\Gamma(k-n+1)}
\right\rangle.
\label{qn-neutral}
\end{equation}

We see (when the moments of $k$ are finite) that $q_2$ 
is much larger than all the other $q_n$ when the size $N$ 
of the population is large, and therefore in the ancestry
of a finite number $n$ of individuals, branches coalesce only by pairs.
Similarly, the probability that two or more pairs of individuals coalesce at
the same generation is negligible.

Let $T_n(g)$ be the age of the most recent common  ancestor of a 
group of $n$ individuals at generation $g$.
As for large~$N$ only coalesences by pairs may occur from one generation to the
previous one, one  has
\begin{equation}
T_n(g+1)   = 
\begin{cases}
\displaystyle
T_n(g) + 1   & \text{with probability}\ \  
1 - {\frac12{n(n-1)}} q_2\vspace{5pt} \\
\displaystyle
T_{n-1}(g) + 1 & \text{with probability}\ \ 
 \frac12{n(n-1)} q_2.
\end{cases}
\end{equation}
In the steady state \cite{SimonDerrida.06}, this implies that
\begin{equation}
\big\langle  T_n^p \big\rangle = \left(1 - \frac{n(n-1)}{2} q_2  \right) 
\big\langle  (1+T_n)^p \big\rangle  + \frac{n(n-1)}{2} q_2 \big\langle 
 (1+T_{n-1})^p \big\rangle 
\label{recursion-neutral}
\end{equation}
and  using the fact that $T_1(g)=0$ and   one gets
\begin{equation}
\label{Tn-neutral}
\langle T_n \rangle =\left( 2 - \frac{2}{n} \right) \frac{1}{q_2}.
\end{equation}
We see that all the times $T_n$ scale like $N$ (since $q_2 \sim N^{-1}$)
and that
\begin{equation}
\frac{\langle T_3 \rangle}{\langle T_2 \rangle} = \frac43, \qquad
\frac{\langle T_4 \rangle}{\langle T_2 \rangle} = \frac32, \qquad
\ldots,\qquad
\frac{\langle T_n \rangle}{\langle T_2 \rangle} = \frac{2(n-1)}{n}.
\end{equation}
One can also calculate from (\ref{recursion-neutral}) higher moments 
of the $T_n$'s or their  generating functions
\begin{equation}
\frac{\langle (T_2)^2 \rangle}{\langle T_2 \rangle^2} = 2, \qquad
\frac{\langle (T_3)^2 \rangle}{\langle T_3 \rangle^2} = \frac{13}{8}.
\end{equation}
These distributions of the $T_n$ as well as their correlations are universal
(in the sense that they do not depend on the 
details of the distribution of the $p_k$'s).



\end{document}